\documentclass[aps,twocolumn,english,prb,showpacs,superscriptaddress]{revtex4-1}
\usepackage[colorlinks=true,urlcolor=blue,citecolor=blue,linkcolor=blue]{hyperref}

\usepackage[T1]{fontenc}
\usepackage[latin9]{inputenc}
\usepackage{amssymb}
\usepackage{graphicx}
\usepackage{amsmath,color}
\usepackage{mathrsfs}
\usepackage{float}
\usepackage{indentfirst}
\usepackage{subfigure}
\usepackage{multirow}
\usepackage{tabu}
\usepackage{booktabs}
\usepackage{txfonts}
\usepackage{amsmath,amssymb,bbm}

\usepackage[euler]{textgreek}

\usepackage{graphicx}
\usepackage{bm}

\newcommand{\Eq}[1]{Eq.~(\ref{#1})}

\begin{document}

\preprint{APS/123-QED}

\title{Klein bottle entropy of compactified boson conformal field theory}%

\author{Wei Tang}
\author{X.~C. Xie}
 \affiliation{International Center for Quantum Materials, School of Physics, Peking University, Beijing 100871, China}
\author{Lei Wang}
 \email{wanglei@iphy.ac.cn}
 \affiliation{Beijing National Lab for Condensed Matter Physics and Institute of Physics, Chinese Academy of Sciences, Beijing 100190, China}
 \affiliation{Songshan Lake Materials Laboratory, Dongguan, Guangdong 523808, China} 
\author{Hong-Hao Tu}
 \email{hong-hao.tu@tu-dresden.de}
 \affiliation{Institute of Theoretical Physics, Technische Universit\"at Dresden, 01062 Dresden, Germany}

\date{\today}

\begin{abstract}
We extend the scope of the Klein bottle entropy, originally introduced by [\href{https://journals.aps.org/prl/abstract/10.1103/PhysRevLett.119.261603}{Tu, Phys.~Rev.~Lett.~119, 261603 (2017)}] in the rational conformal field theory (CFT), to the compactified boson CFT, which are relevant to the studies of Luttinger liquids. We first review the Klein bottle entropy in rational CFT and discuss details of how to extract the Klein bottle entropy from lattice models using the example of the transverse field Ising model.
We then go beyond the scope of rational CFT and study the Klein bottle entropy $\ln g$ in the compactified boson CFT, which turns out to have a straightforward relation to the compactification radius $R$, $\ln g = \ln R$.
This relation indicates a convenient and efficient method to extract the Luttinger parameter from lattice model calculations.
Our numerical results on the Klein bottle entropy in the spin-$1/2$ XXZ chain show excellent agreement with the CFT predictions, up to some small deviations near the isotropic point, which we attribute to the marginally irrelevant terms.
For the $S = 1$ XXZ chain that cannot be exactly solved, our numerical results provide an accurate numerical determination of the Luttinger parameter in this model.
\end{abstract}

\maketitle



\section{Introduction}

Conformal field theory{~\cite{francesco_conformal_1999}} (CFT) has become the center of much interest during the past decades. Due to its powerful nature in two dimensions, it has been widely applied to study the universal behavior at the critical points of two-dimensional statistical systems and one-dimensional quantum systems, where the correlation length of the system diverges.
Notable applications include the classification of universality classes{~\cite{francesco_conformal_1999}}, the gapless edge modes of fractional quantum Hall systems {~\cite{moore_nonabelions_1991}}, the entanglement entropy{~\cite{holzhey_geometric_1994,vidal_entanglement_2003,calabrese_entanglement_2004, kitaev_topological_2006, levin_detecting_2006, fendley2007topological, fradkin_entanglement_2006}}, and the Kondo problem{~\cite{affleck_conformal_1995}}.

Recently, the entropy correction on a Klein bottle is proposed to be a universal characterization of the critical systems described by CFT, which is called the Klein bottle entropy{~\cite{tu_universal_2017}}.
The path integral on a Klein bottle is achieved by swapping the left movers and the right movers via a reflection operator defined on the CFT Hilbert space. The operation swaps the world line and glues them back via taking the trace in the path integral.
This result is soon generalized to other nonorientable manifolds, such as the M\"obius strip{~\cite{chen_conformal_2017}} and the real projective plane{~\cite{wang_logarithmic_2018}}.
In conformal critical systems, the Klein bottle entropy is a universal value which only depends on the type of the CFT,
and thus it can be applied to characterize the underlying CFT description of the system.
The Klein bottle entropy can also be used to accurately pinpoint quantum critical points, even those without local order parameters{~\cite{chen_conformal_2017}}.
In lattice models, the Klein bottle entropy can be efficiently calculated using prevailing numerical algorithms,
such as quantum Monte Carlo{~\cite{tang_universal_2017}} and thermal tensor network methods{~\cite{chen_conformal_2017,wang_logarithmic_2018}}.
In these lattice model calculations, however, it can be a subtle issue to identify the correct lattice operation which exactly exchange the left movers and right movers in the CFT level. Therefore, one should carry out a careful analysis in order to confirm that the results of the lattice simulation match the CFT predictions.

The initial work Ref.~{\onlinecite{tu_universal_2017}} of the Klein bottle entropy only concentrates on the rational CFT (RCFT), whose space of states can be decomposed into a finite number of representations of the Virasoro algebra or other extended chiral algebra (such as the Kac-Moody algebra). It is interesting to investigate how the Klein bottle entropy extends to a broader class of CFTs. 

The main focus of the present work is to study the Klein bottle entropy in another notable category of CFT, the free boson theory compactified on a circle, which includes both rational and nonrational CFTs.
The compactified boson CFTs all have the identical central charge $c = 1$ and are characterized by the compactification radius $R$.
In condensed matter physics, the compactified boson CFT plays an important role through its connection to the Luttinger liquid theory, which is a remarkably successful and powerful framework describing the low-energy physics of one-dimensional critical systems{~\cite{ludwig_methods_1995,giamarchi_quantum_2003}}.
The Luttinger liquids are relevant to various experimental systems, such as carbon nanotubes{~\cite{bockrath_luttinger-liquid_1999,yao_carbon_1999,ishii_direct_2003,gao_evidence_2004}} , semiconductor wires{~\cite{yacoby_nonuniversal_1996,levy_luttinger-liquid_2006}}, and highly tunable ultracold atomic gases~\cite{kinoshita_observation_2004, clement_exploring_2009, haller_pinning_2010}.

The Luttinger liquid theory is fully characterized by two parameters, the sound velocity $v$ and the Luttinger parameter $K$.
The value of the Luttinger parameter $K$ has a direct relation with the compactification radius $R$ of the free boson CFT\footnote{the specific form of this relation depends on the choice of the dual field, as will be discussed in Sec.~\ref{sec:determineR}}. 
However, the value of the Luttinger parameter cannot be reliably determined in field theory calculations, and one usually has to resort to the microscopic models to obtain its value which is usually still a nontrival task{~\cite{giamarchi_quantum_2003,song_general_2010,song_bipartite_2012,dalmonte_critical_2012,dalmonte_estimating_2011,lauchli_operator_2013,alcaraz_in_preparation_2018}}. It is highly desired to have a direct determination of the Luttinger parameter without any finite-size scaling or fitting procedure.

In this paper, by studying the Klein bottle entropy of the compactified boson CFT, we discover a simple relation between the Klein bottle entropy $\ln g$ and the compactification radius $R$, $\ln g = \ln R$.
This simple relation suggests an efficient and accurate method to extract the Luttinger parameter from lattice models in a straightforward manner.
To verify this relation numerically, we perform quantum Monte Carlo (QMC) calculations in the $S = 1 / 2$ XXZ chain, whose Luttinger parameter (and thus the compactification radius) can be exactly obtained from the Bethe ansatz solution{~\cite{giamarchi_quantum_2003}}.
As an application, we present numerical results in the $S = 1$ XXZ chain, which cannot be exactly solved, and our results serve as an accurate numerical determination of the Luttinger parameter in this model.

This paper is organized as follows.
In Sec.~\ref{sec:RCFT}, we review the main results of Ref.~{\onlinecite{tu_universal_2017}}, by introducing the definition of the Klein bottle entropy and deriving its RCFT prediction.
We also show how to extract the Klein bottle entropy from lattice models by discussing the transverse field Ising model (TFIM) in detail.
In Sec.~\ref{sec:CBCFT}, we present the main result of this work, the prediction of the Klein bottle entropy in the compactified boson CFT, and perform the numerical calculations in the XXZ model with spin $S = 1 / 2$ and $S = 1$.
Sec.~\ref{sec:summary} summarizes the results.
In Appendix \ref{sec:extendedMCmethod}, we discuss the details of the extended ensemble Monte Carlo method in the XXZ chain, and in Appendix \ref{sec:JWsolTFIMXY}, we present the exact solutions of the Klein bottle entropy for the two solvable lattice models, the TFIM and the XY chain.

\section{The Klein bottle entropy of rational CFT}\label{sec:RCFT}

In this section, we mainly review the results of Ref.~{\onlinecite{tu_universal_2017}}, which concentrates on RCFT.
We first review the definition of the Klein bottle entropy and derive the prediction of its value in RCFT\footnote{Here we only consider unitary CFTs. For non-unitary CFTs, the RCFT prediction of the Klein bottle entropy needs to be modified.}.
In order to show how to extract the Klein bottle entropy from lattice models, we discuss the example of TFIM in detail.
We show that, in lattice models, the effect of the reflection operator defined in the context of the CFT may be actualized by a bond-centered lattice reflection, but whether this lattice reflection would lead to the Klein bottle entropy predicted by the CFT is not obvious, and one usually needs to perform a careful analysis to confirm this.

\subsection{CFT prediction}

Let us consider a (1+1)-dimensional quantum chain with length $L$ and periodic boundary condition. At inverse temperature $\beta=1/T$ (the Boltzmann constant $k_{\text{B}}$ is set to be 1), its partition function can be written as a path integral defined on a torus with size $L\times \beta$. When the system is critical and its low-energy effective theory is a CFT, the partition function becomes the torus partition function $Z^{\mathcal{T}}$ of the CFT ($\mathcal{T}$ stands for torus){~\cite{francesco_conformal_1999}}
\begin{equation}
  Z^{\mathcal{T}} = \mathrm{Tr}_{\mathcal{H} \otimes \overline{\mathcal{H}}}
  (q^{L_0 - c / 24} \bar{q}^{\bar{L}_0 - c / 24}),
\end{equation}
where $\mathcal{H} \otimes \overline{\mathcal{H}}$ represents the tensor-product Hilbert space of the holomorphic sector $\mathcal{H}$ and the antiholomorphic sector $\overline{\mathcal{H}}$ of the CFT.
$L_0$ and $\bar{L}_0$ are the zeroth-level holomorphic and antiholomorphic Virasoro generators.
$c$ is the central charge. 
$q = \mathrm{e}^{2 \pi \mathrm{i} \tau}$ with $\tau = \mathrm{i} v \beta / L $, where $v$ is the velocity of the CFT,
and $\bar{q}$ is the complex conjugate of $q$.

In the CFT, the Klein bottle partition function $Z^{\mathcal{K}}$ ($\mathcal{K}$ denotes the Klein bottle) is defined by~\cite{blumenhagen_introduction_2009}
\begin{equation}
  Z^{\mathcal{K}}  = \mathrm{Tr}_{\mathcal{H} \otimes
  \overline{\mathcal{H}}} (\Omega q^{L_0 - c / 24} \bar{q}^{\bar{L}_0 - c /
  24}), \label{klein-part}
\end{equation}
where an extra operator $\Omega$ is inserted, which effectively interchanges the holomorphic and antiholomorphic sectors, $\Omega | \alpha, \bar{\mu} \rangle = | \mu, \bar{\alpha} \rangle$, i.e., interchanges the left and right movers.
As a result, only the left-right symmetric states $| \alpha, \bar{\alpha} \rangle$ have contributions to the Klein bottle partition function.
One can then write $Z^{\mathcal{K}}$ as
\begin{equation}
  Z^{\mathcal{K}}  = \mathrm{Tr}_{\mathrm{sym}} (q^{2 (L_0 - c / 24)}), \label{klein-part2}
\end{equation}
where the subscript ``sym'' indicates that the trace in (\ref{klein-part2}) is taken over the left-right symmetric states $|\alpha, \bar{\alpha} \rangle$ in $\mathcal{H} \otimes \overline{\mathcal{H}}$.

For rational CFTs, the Hilbert space can be organized into a finite number of conformal towers, each of which is formed by a primary state and its descendant states.
In such CFTs, the torus partition function is given by
\begin{equation}
  Z^{\mathcal{T}} = \sum_{a, b} \chi_a (q) M_{a, b} \bar{\chi}_b (\bar{q}) .
\end{equation}
Here $\chi_a (q) = \mathrm{Tr}_a (q^{L_0 - c / 24})$ is called a character, where $a$ labels the primary state of the conformal tower, and the trace is over the conformal tower of states.
$\bar{\chi}$ is the antiholomorphic correspondence of $\chi$. 
$M_{a, b}$ represents the element of the $M$-matrix, which are non-negative integers representing the number of primary states $(a, \bar{b})$ in the Hilbert space $\mathcal{H} \otimes \overline{\mathcal{H}}$.
On the other hand, according to Eq.~\eqref{klein-part2}, one can write $Z^{\mathcal{K}}$ as
\begin{equation}
  Z^{\mathcal{K}} = \sum_a M_{a, a} \chi_a (q^2).
\end{equation}

In the limit $L \gg v \beta$, the partition functions can be evaluated by using the modular transformation of the characters, i.e., $\chi_a (q) = \sum_b S_{a b} \chi_b (q')$, where $q = \mathrm{e}^{- 2 \pi \frac{v \beta}{L}}$ and $q' = \mathrm{e}^{- 2 \pi \frac{L}{v \beta}}$, and $S_{a b}$ is the element of the modular S matrix{~\cite{francesco_conformal_1999}}. When $L \gg v \beta$, $q' \rightarrow 0$, then in the character $\chi_a (q')$ the primary state $a$ dominates, so $\chi_a(q') \approx (q')^{h_a - c / 24}$, where $h_a$ is the conformal weight of the primary field $a$.
Furthermore, among all primary fields, the identity field with conformal weight $h_I = 0$ dominates over other primary fields with $h_a > 0$.

Based on the discussion above, for the torus partition function, due to the modular invariance of the partition function $M = S^{\dagger} M S$ and uniqueness of the identity field $M_{I, I} = 1$, one obtains
\begin{equation}
  Z^{\mathcal{T}} = \sum_{a, b} \chi_a (q') M_{a, b} \bar{\chi}_b
  (q') \approx | \chi_I (q') |^2 = \mathrm{e}^{\frac{\pi c L}{6
  v \beta}} .
\end{equation}
Meanwhile, for the Klein bottle partition function, since $\chi_a (q^2) = \sum_b S_{a b} \chi_b (q'^{1 / 2}) \approx S_{a I} \chi_I (q'^{1 / 2})$, we have
\begin{equation}
  Z^{\mathcal{K}} \approx \sum_a M_{a, a} S_{a I} \chi_I (q'^{1 / 2}) = g
  \mathrm{e}^{\frac{\pi c L}{24 v \beta}},
\end{equation}
where we have introduced
\begin{equation}
  g = \sum_a M_{a, a} S_{a I} = \sum_a \frac{M_{a, a} d_a}{\mathcal{D}}.
  \label{eq:gforrationalCFT}
\end{equation}
Here $d_a$ is the quantum dimension of the primary field $a$ and $\mathcal{D} = \sqrt{\sum_a d^2_a}$ is the total quantum dimension, which satisfy $S_{aI} = d_a/\mathcal{D}$.

In lattice models, besides the pure CFT predictions, one has to take into account the nonuniversal free energy terms,
\begin{eqnarray}
  \ln Z^{\mathcal{T}} & \approx & - f_0 \beta L + \frac{\pi c}{6 v \beta} L, \label{toruslatticeF}
  \\
  \ln Z^{\mathcal{K}} & \approx & - f_0 \beta L + \frac{\pi c}{24 v \beta} L +
  \ln g, \label{kleinlatticeF}
\end{eqnarray}
where $f_0$ represents the bulk free energy density and $\ln g$ is the Klein bottle entropy, which is universal and only depends on the quantum dimensions of the primary fields [see Eq.~(\ref{eq:gforrationalCFT})].
We note that Eq.~\eqref{toruslatticeF} is the seminal result obtained in Refs.~\onlinecite{affleck_universal_1986, blote_conformal_1986},
and Eq.~\eqref{kleinlatticeF} is the central result of Ref.~\onlinecite{tu_universal_2017}.
Actually, one can cancel the nonuniversal terms in (\ref{toruslatticeF}) and (\ref{kleinlatticeF}) and extract the Klein bottle entropy $\ln g$ by calculating the following partition function ratio:
\begin{equation}
  \ln g = \ln \frac{Z^{\mathcal{K}} \left( 2 L, \frac{\beta}{2}
  \right)}{Z^{\mathcal{T}} (L, \beta)} . \label{eq:Kleinbottleexpression}
\end{equation}
We emphasize that the Klein bottle entropy is universal, which is unchanged even in the zero-temperature limit $\beta\rightarrow\infty$, as long as the thermodynamic limit $L\rightarrow\infty$ is taken first.
In this regard, the Klein bottle entropy reflects the ground-state properties of the system.
Therefore, Eq.~\eqref{eq:Kleinbottleexpression} allows one to extract the ground-state properties directly from thermal systems, without any fitting procedure.

\subsection{Transverse field Ising model}
\label{subsec:Ising}

As a concrete example, we consider the spin-1/2 critical Ising chain,
\begin{equation}
  H_{\mathrm{Ising}} = - \sum_{i = 1}^L (S_i^x S_{i + 1}^x + \frac{1}{2} S_i^z),
   \label{eq:IsingHamil}
\end{equation}
which is well known to be described by the Ising CFT. Here we have imposed periodic boundary condition, i.e., $S^{\nu}_{L + 1} = S_1^{\nu}\, (\nu = x, y, z)$.
For simplicity, we only consider the case of even $L$.

The spin-1/2 critical Ising chain can be transformed into a spinless fermion model via the Jordan-Wigner transformation, $S_i^z = f_i^{\dagger} f_i - \frac{1}{2}$ and $S_i^x = \frac{1}{2} ( f_i^{\dagger}+f_i) \mathrm{e}^{\mathrm{i} \pi \sum_{l < i} n_l} $. The Hamiltonian (\ref{eq:IsingHamil}) is then fermionized as
\begin{eqnarray}
  H & = & - \frac{1}{4} \sum_{i = 1}^{L - 1} (f_i^{\dagger} - f_i) (f_{i +
  1}^{\dagger} + f_{i + 1}) - \frac{1}{4} \sum_{i = 1}^L (2 f_i^{\dagger} f_i
  - 1) \nonumber\\
  &  & + \frac{1}{4} Q (f_L^{\dagger} - f_L) (f_1^{\dagger} + f_1),
  \label{eq:IsingfermionizedHamil}
\end{eqnarray}
where the fermion parity $Q = \mathrm{e}^{i \pi \sum_{l = 1}^L n_l} = \pm 1$ is a conserved quantity in this model.

The Hilbert space splits into two sectors with definite fermion parity in each sector. The two sectors, following the CFT convention, are called Neveu-Schwarz and Ramond sectors, respectively{~\cite{francesco_conformal_1999}}.
In the Neveu-Schwarz sector, the fermion parity is even ($Q = 1$), with allowed lattice momenta $k = \pm \frac{\pi}{L}, \pm \frac{3 \pi}{L}, \ldots, \pm \frac{(L - 1) \pi}{L}$.
In the Ramond sector, the fermion parity is odd ($Q = - 1$), with allowed lattice momenta $k = 0, \pm \frac{2 \pi}{L}, \pm \frac{4 \pi}{L}, \ldots, \pm \frac{(L - 2) \pi}{L}, \pi$.
In the two sectors, the Hamiltonian (\ref{eq:IsingfermionizedHamil}) takes the same form
\begin{equation}
  H_{\pm} = - \frac{1}{4} \sum_{i = 1}^L (f_i^{\dagger} - f_i) (f_{i +
  1}^{\dagger} + f_{i + 1}) - \frac{1}{4} \sum_{i = 1}^L (2 f_i^{\dagger} f_i
  - 1),
\end{equation}
but with different boundary conditions for fermions. In the Neveu-Schwarz (Ramond) sector, the fermions have antiperiodic (periodic) boundary condition $f_{L + 1} = - f_1$ ($f_{L + 1} = f_1$).

\subsubsection{Bond-centered lattice reflection} \label{sec:latticereflection}

For lattice models, one needs to find an operator defined on the lattice, which effectively interchanges the left and right movers when acting on the states of the system.
As indicated in Ref.~{\onlinecite{tu_universal_2017}}, the following bond-centered lattice reflection operator $P$ serves as a natural candidate:
\begin{equation}
  P|s_1, s_2, \ldots, s_{L - 1}, s_L \rangle = |s_L, s_{L - 1}, \ldots, s_2,
  s_1 \rangle,
\end{equation}
where $s_i$ represents the spin state at site $i$.

Next, we need to work out the action of the reflection operator $P$ in the fermionic basis.
From $P S_i^{\nu} P^{- 1} = S^{\nu}_{L+1-i}$, one obtains $P f_i^{\dagger} P^{- 1} = f_{L - i + 1}^{\dagger} Q$ with the help of the Jordan-Wigner transformation, and in the momentum space
\begin{equation}
  P f_k^{\dagger} P^{- 1} = \mathrm{e}^{\mathrm{i} (L + 1) k} f_{-
  k}^{\dagger} Q. \label{latticePMomentum}
\end{equation}
According to Eq.~{\eqref{latticePMomentum}}, up to a phase factor, the lattice reflection reflects a fermion mode of momentum $k$ to a fermion mode of momentum $- k$.
As a result, one can infer that only a few states which are composed of ``fermion pairs'' like $f^{\dagger}_{- k} f^{\dagger}_k$ (except $k = 0, \pi$, since the corresponding fermion mode is reflected to itself, up to a phase factor) will contribute to the Klein bottle partition function, while most other states in the Hilbert space are orthogonal to their reflection partners.
To construct the states that are invariant under lattice reflection, one can consider a state $| \psi \rangle$ that is invariant under the lattice reflection, i.e., $P| \psi \rangle = | \psi \rangle$, and create fermion modes on top of this state.
According to Eq.~{\eqref{latticePMomentum}}, one can easily see that
\begin{eqnarray}
  P f^{\dagger}_{- k} f^{\dagger}_k | \psi \rangle & = &
  (\mathrm{e}^{-\mathrm{i} (L + 1) k} f_k^{\dagger} Q) (\mathrm{e}^{
  \mathrm{i} (L + 1) k} f_{-k}^{\dagger} Q) P| \psi \rangle \nonumber\\
  & = & f_k^{\dagger} Q f_{- k}^{\dagger} Q | \psi \rangle \nonumber\\
  & = & f^{\dagger}_{- k} f^{\dagger}_k | \psi \rangle.
\end{eqnarray}
On the other hand, we note that the vacuum state $|0 \rangle$ corresponds to the spin fully polarized state and it is invariant under the lattice reflection, i.e., $P| 0 \rangle = |0 \rangle$. What is more,
\begin{eqnarray}
  P f_{k = 0}^{\dagger} |0 \rangle & = & f_{k = 0}^{\dagger} |0 \rangle, \\
  P f_{k = \pi}^{\dagger} |0 \rangle & = & - f_{k = \pi}^{\dagger} |0 \rangle
  .
\end{eqnarray}
Therefore, the states generated by creating fermion pairs like $f^{\dagger}_k f^{\dagger}_{- k}$ on top of $|0 \rangle$ or $f_{k = 0}^{\dagger} |0 \rangle$ are invariant under lattice reflection, and these states have total momentum $k_{\mathrm{tot}} = 0$.
Meanwhile, states generated by creating fermion pairs on top of $f_{k = \pi}^{\dagger} |0 \rangle$ have total momentum $k_{\mathrm{tot}} = \pi$, and these states are invariant under lattice reflection up to a sign factor $- 1$.
States in the other forms are all orthogonal to their reflection partners.

In the lattice models described by RCFT, in order to verify whether the lattice reflection will lead to the Klein bottle entropy predicted by CFT, we need to identify the primary states and investigate their behavior under the lattice reflection.
Only the quantum dimensions of the primary states that are invariant under the lattice reflection can be counted in the summation of Eq.~\eqref{eq:gforrationalCFT}.
We also note that sometimes there may exist primary states that are invariant under lattice reflection up to a sign factor $-1$.
In such cases, when calculating the Klein bottle entropy, one needs to add an additional minus sign before the corresponding quantum dimension in Eq.~\eqref{eq:gforrationalCFT}~\cite{chen_conformal_2017}.

\subsubsection{Identification of the primary states of Ising CFT}

As discussed above, in order to calculate the Klein bottle entropy of the critical Ising chain using the RCFT prediction Eq.~\eqref{eq:gforrationalCFT}, we need to identify the primary states in the fermionic picture and analyze their behavior under the lattice reflection.
We also obtain the energy spectrum of the critical Ising chain of size $L=14$ by means of exact diagonalization calculations, and then, as a separate check, we identify the primary states obtained from the fermionic picture in this energy spectrum.

In the Neveu-Schwarz sector, $Q = 1$, $k = \pm \frac{\pi}{L}, \pm \frac{3
\pi}{L}, \ldots, \pm \frac{(L - 1) \pi}{L}$. By a Fourier transformation, the
Hamiltonian becomes
\begin{eqnarray}
  H_+ & = & \frac{1}{4} \sum_{k > 0} (f_k^{\dagger}, f_{- k}) \left(
  \begin{array}{cc}
    - 2 \mathrm{cos} k - 2 & - 2 \mathrm{i} \mathrm{sin} k\\
    2 \mathrm{i} \mathrm{sin} k & 2 \mathrm{cos} k + 2
  \end{array} \right) \left( \begin{array}{c}
    f_k\\
    f_{- k}^{\dagger}
  \end{array} \right) \nonumber\\
  &  & + \frac{1}{4} \sum_{k > 0} \left( - 2 \mathrm{cos} k - 2 \right) +
  \frac{L}{4} .  \label{eq:IsingNShamilt}
\end{eqnarray}
By diagonalizing the matrix in Eq.~\eqref{eq:IsingNShamilt}, one can obtain the dispersion relation $\varepsilon_k = \pm E_k $, where we have introduced $E_k = \mathrm{cos} (k / 2)$.
We then write $H_+$ as
\begin{equation}
  H_+ = \sum_{k > 0} E_k (\alpha_k^{\dagger} \alpha_k + \beta_k
  \beta_k^{\dagger}) - \sum_{k > 0} E_k + \frac{1}{4} \sum_{k > 0} \left( - 2
  \mathrm{cos} k - 2 \right) + \frac{L}{4},
\end{equation}
where $\alpha_k = f_k \sin \frac{k}{4} + f_{- k}^{\dagger} \cos \frac{k}{4}$,\, $\beta_k = \mathrm{i} f_k \cos \frac{k}{4} - \mathrm{i} f_{- k}^{\dagger} \sin \frac{k}{4}$. 
We introduce 
\begin{equation}
  \gamma_k = \left\{ \begin{array}{ll}
    \alpha_k & \text{for } k > 0\\
    \beta_{- k}^{\dagger} & \text{for } k < 0
  \end{array} \right. 
\end{equation}
and note the ground-state energy is
\begin{equation}
  E_{\mathrm{gs}} = \frac{L}{4} - \frac{1}{4} \sum_{k > 0} \left( E_k + 2
  \mathrm{cos} k + 2 \right) = - \frac{1}{2 \sin \left( \frac{\pi}{2 L}
  \right)} . \label{eq:isinggsE}
\end{equation}
so the Hamiltonian becomes
\begin{equation}
  H_+ = \sum_{k \neq 0} E_k \gamma_k^{\dagger} \gamma_k + E_{\mathrm{gs}} .
\end{equation}
The low-energy states in the Neveu-Schwarz sector are generated by creating fermion modes $\gamma_k^{\dagger}$'s on top of the ground state $| \mathrm{gs} \rangle$.

\begin{figure}[!htb]
  \resizebox{8.5cm}{!}{\includegraphics{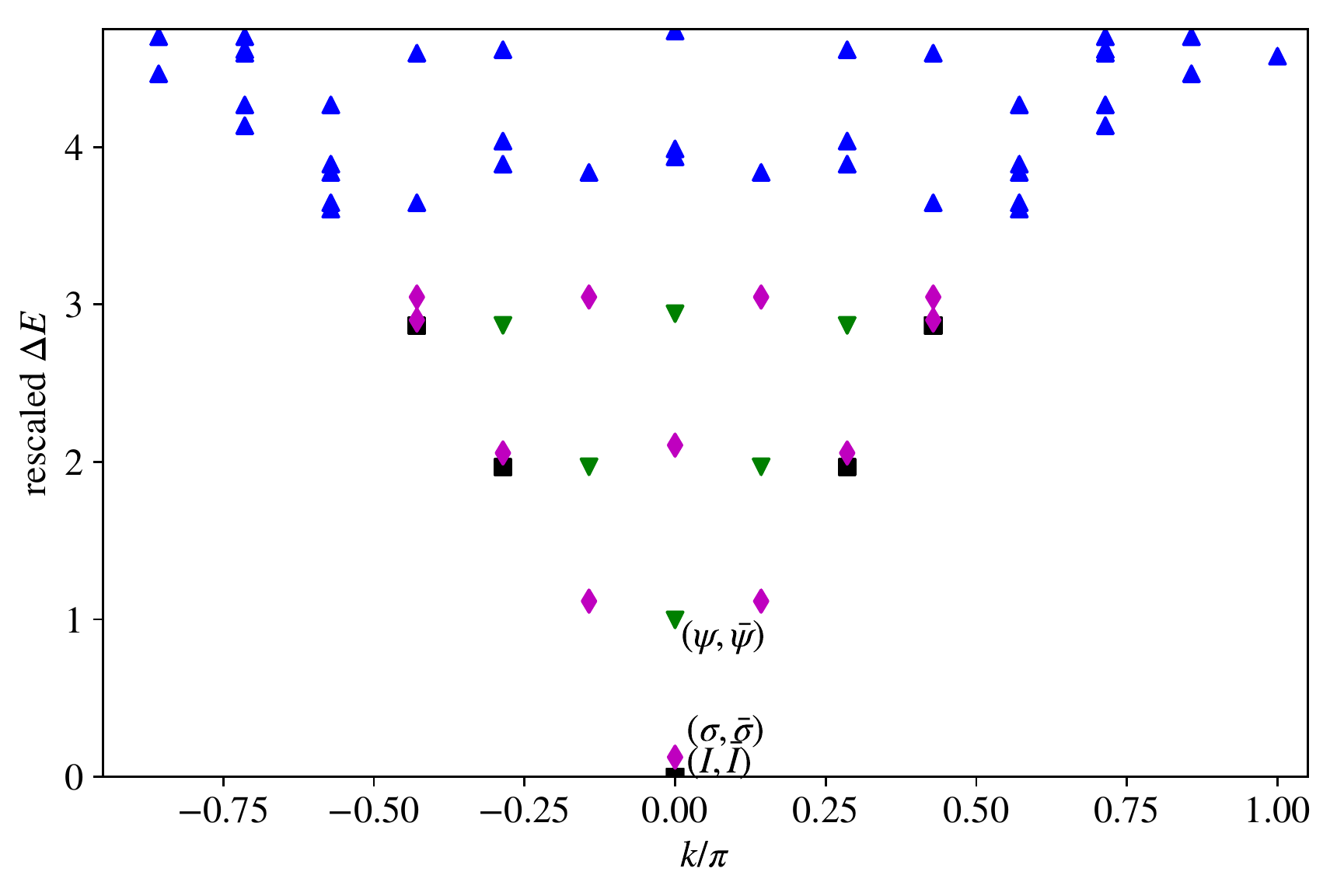}}
  \caption{The low-energy spectrum of the critical Ising chain of size $L = 14$.
  The primary states are marked by $(I, \bar{I})$, $(\sigma, \bar{\sigma})$ and $(\psi, \bar{\psi})$ in the spectrum. The energy differences $\Delta E$'s have been rescaled according to the scaling dimensions of the primary fields.
  The conformal towers of descendant states are marked by the same color with the corresponding primary states, while the unidentified states are marked as blue. \ \label{fig:ising-spectrum}}
\end{figure}

In order to determine whether the ground state is invariant under lattice reflection, we have to derive the ground-state wave function.
To do this, we write the Hamiltonian in the subspace of $|0 \rangle$ and $f_k^{\dagger} f_{-k}^{\dagger} |0 \rangle$,
\begin{equation}
  H_+^k = \left( \begin{array}{cc}
    0 & - \frac{\mathrm{i}}{2} \mathrm{sin} k\\
    \frac{\mathrm{i}}{2}  \mathrm{sin} k & - \cos k - 1
  \end{array} \right) .
\end{equation}
By diagonalizing this matrix, we obtain
\begin{equation}
  | \mathrm{gs} \rangle = \prod_{k > 0} (u_k + v_k f_k^{\dagger} f_{-
  k}^{\dagger}) |0 \rangle,
\end{equation}
where $u_k = \mathrm{i} \sin (k / 4),\, v_k = \cos (k / 4)$.
Apparently, the ground state has total momentum $0$ and it is invariant under the lattice reflection.
The ground state corresponds to the primary field $(I,\bar{I})$ in the Ising CFT, which has the conformal weight $(h_I, \bar{h}_{\bar{I}}) = (0, 0)$.
This state is labeled by $(I, \bar{I})$ in the energy spectrum Fig.~\ref{fig:ising-spectrum}.

The lowest excitation state in the Neveu-Schwarz sector is
\begin{equation}
  | \psi, \bar{\psi} \rangle = \gamma_{k=(1-2/L)\pi}^{\dagger} \gamma_{k=-(1-2/L)\pi}^{\dagger} |
  \mathrm{gs} \rangle .
\end{equation}

Note that for $k > 0$
\begin{equation}
  P \gamma_k^{\dagger} P^{- 1} = P \alpha_k^{\dagger} P^{- 1} = - \mathrm{i}
  \mathrm{e}^{\mathrm{i} k} \beta_k Q = - \mathrm{i} \mathrm{e}^{\mathrm{i} k}
  \gamma_{- k}^{\dagger} Q.
\end{equation}
One can infer that $| \psi, \bar{\psi} \rangle$ is invariant under the lattice
reflection. This state has momentum 0. The energy of this state is
\begin{equation}
  E_{(\psi, \bar{\psi})} = E_{\mathrm{gs}} + 2 \cos \left( \frac{L - 2}{2 L}
  \pi \right) = - \frac{1}{2 \sin \left( \frac{\pi}{2 L} \right)} + 2 \sin
  \left( \frac{\pi}{2 L} \right) .
\end{equation}
This state corresponds to the primary field $(\psi, \bar{\psi})$ with conformal weight $(h_{\psi}, \bar{h}_{\bar{\psi}}) = (1 / 2, 1 / 2)$.
We label it by $(\psi, \bar{\psi})$ in the energy spectrum Fig.~\ref{fig:ising-spectrum}.

In the Ramond sector, $Q = - 1$, $k = 0, \pm \frac{2 \pi}{L}, \pm \frac{4 \pi}{L}, \ldots, \pm \frac{(L - 2) \pi}{L}, \pi$.
Similarly, the Hamiltonian in this sector can be written as
\begin{eqnarray}
  H_- & = & \frac{1}{4} \sum_{0 < k < \pi} (f_k^{\dagger}, f_{- k}) \left(
  \begin{array}{cc}
    - 2 \mathrm{cos} k - 2 & - 2 \mathrm{i} \mathrm{sin} k\\
    2 \mathrm{i} \mathrm{sin} k & 2 \mathrm{cos} k + 2
  \end{array} \right) \left( \begin{array}{c}
    f_k\\
    f_{- k}^{\dagger}
  \end{array} \right) \nonumber\\
  &  & - f^{\dagger}_{k = 0} f_{k = 0} + \frac{1}{4} \sum_{0 < k < \pi}
  \left( - 2 \mathrm{cos} k - 2 \right) + \frac{L}{4} ,
  \label{eq:TRIMHinRamond}
\end{eqnarray}
where the fermionic mode $f^{\dag}_{k=\pi}$ does not show up since its single-particle energy is zero. 
The energy dispersion $\varepsilon_k = \pm E_k = \pm \cos (k / 2)$.
The energy of the lowest state in the Ramond sector is
\begin{equation}
  E_{(\sigma, \bar{\sigma})} = - \frac{1}{2} - \frac{1}{2} \sum_{0 < k < \pi}
  \left( \cos k + 2 \cos \frac{k}{2} \right) = - \frac{1}{2} \cot \left(
  \frac{\pi}{2 L} \right) . \label{eq:isingsigmaE}
\end{equation}
Note that since $\varepsilon_{k=0} = -1$ [see Eq.~\eqref{eq:TRIMHinRamond}], the $k=0$ mode is occupied in $|\sigma,\bar{\sigma}\rangle$, and the $k=\pi$ state is therefore unoccupied due to the odd fermion parity constraint. 
Similarly as above, one can obtain the wave function for this state
\begin{equation}
  |\sigma,\bar{\sigma}\rangle = f^{\dag}_{k=0} \prod_{0<k<\pi}(u_k + v_k f^{\dag}_{k} f^{\dag}_{-k}) |0\rangle,
\end{equation}
which is invariant under the lattice reflection.
This state has total momentum $k_{\mathrm{tot}} = 0$ and corresponds to the primary field $(\sigma, \bar{\sigma})$ in the Ising CFT, with conformal weight $(h_{\sigma}, \bar{h}_{\bar{\sigma}}) = (1 / 16, 1 / 16)$.
We label it by $(\sigma, \bar{\sigma})$ in the energy spectrum Fig.~\ref{fig:ising-spectrum}.

From the discussions above, we see that the three primary states of Ising CFT all have total momentum $k_{\mathrm{tot}} = 0$, and they are all invariant under the lattice reflection.
As a result, the corresponding three primary fields all contribute to the Klein bottle partition function.
The quantum dimensions of the primary fields $I$, $\psi$, and $\sigma$ are, respectively 1, 1, and $\sqrt{2}$, and the total quantum dimension $\mathcal{D}=2$.
From Eq.~\eqref{eq:gforrationalCFT}, one can then obtain
\begin{equation}
  g_{\mathrm{Ising}} = \frac{2 + \sqrt{2}}{2} ,
\end{equation}
where we have used $M_{a,a}=1 \, \forall a$ in the Ising CFT.

As a consistency check, one can compute the Klein bottle entropy for the critical Ising chain analytically and compare with the CFT prediction.
As shown in Ref.~{\onlinecite{tu_universal_2017}}, the CFT prediction and the exact solution are consistent with each other.
The details of the exact solution are presented in Appendix \ref{subsec:exacttoising}.

\section{The Klein bottle entropy of the compactified boson CFT}\label{sec:CBCFT}

In this section, we extend the results on Klein bottle entropy to compactified boson CFT, which contains both rational and nonrational CFTs.
As a central result of this work, we present the Klein bottle entropy of the compactified boson CFT, which provides direct access to the compactification radius $R$. 
This result provides a practical numerical method to extract the Luttinger parameter of lattice models, due to the direct relation between the compactification radius and the Luttinger parameter.
As concrete examples, we first discuss the spin-$1 / 2$ XY chain in detail, which can be analyzed from both the rational $U(1)_4$ CFT and the compactified boson CFT perspectives. Next, we extend our discussion to the XXZ chain with $S=1/2$ and $S=1$ and numerically calculate the Klein bottle entropy in the critical phases of these models.

\subsection{CFT prediction}

In the free boson CFT, the descendant states in the Hilbert space are obtained by acting $j_{- k}$ and $\bar{j}_{- k}$ ($k > 0$) on the highest weight states $| \alpha \rangle$ as{~\cite{blumenhagen_introduction_2009}}
\begin{equation}
  j_{- 1}^{n_1} j_{- 2}^{n_2} \ldots \bar{j}_{- 1}^{m_1} \bar{j}_{- 2}^{m_2}
  \ldots | \alpha \rangle \text{ with } m_k, n_k \geqslant 0,
\end{equation}
where $j_{- k}$ ($\bar{j}_{- k}$) is the Laurent mode of the chiral current $j (z) = \mathrm{i} \partial_z \phi (z, \bar{z})$ (antichiral current $\bar{j} (\bar{z}) = \mathrm{i} \partial_{\bar{z}} \phi (z, \bar{z})$) with $\phi (z, \bar{z})$ being the free boson field.
For $k > 0$, $j_{- k}$ ($\bar{j}_{- k}$) plays the role of creation operator of the excitations in the holomorphic (antiholomorphic) sector, and $j_k$ ($\bar{j}_k$) is the annihilation operator of the excitations in the holomorphic (antiholomorphic) sector.
Highest weight states $| \alpha \rangle$ are those states which are annihilated by all annihilation operators, i.e., $j_k | \alpha \rangle = 0$, $\bar{j}_k | \alpha \rangle = 0$ $\forall k > 0$.

When the free boson is compactified on a circle, the highest weight states can be represented as $|n, m \rangle$, where $n, m \in \mathbbm{Z}$.
These states are eigenstates of $j_0$ and $\bar{j}_0$,
\begin{eqnarray}
  j_0 |n, m \rangle & = & \left( \frac{n}{R} + \frac{R m}{2} \right) |n, m
  \rangle,  \label{eq:j0eigen}\\
  \bar{j}_0 |n, m \rangle & = & \left( \frac{n}{R} - \frac{R m}{2} \right) |n,
  m \rangle,  \label{eq:j0bareigen}
\end{eqnarray}
where $R$ is the compactification radius. Here $n$ corresponds to the center of mass momentum, which is quantized due to the existence of the compactification radius $R$.
Meanwhile, $m$ is the winding number of the bosonic field, $\phi (x + L, t) \equiv \phi (x, t) + 2 \pi m R$.

To evaluate the Klein bottle partition function \eqref{klein-part} for compactified boson CFT, one needs to find all the states that are invariant under the reflection operator $\Omega$.
The operator $\Omega$ effectively interchanges the holomorphic and antiholomorphic sectors, or more concretely, in the present case,
\begin{equation}
  \Omega^{-1} j_k \Omega = \bar{j}_k, \; k \in \mathbbm{Z}.
\end{equation}
To determine the reflected state of the highest weight state $|n, m \rangle$, one can act $j_0$ on $\Omega |n, m \rangle$ and get $j_0 \Omega |n, m \rangle = \Omega (\Omega^{- 1} j_0 \Omega) |n, m \rangle = \Omega \bar{j}_0 |n, m \rangle = \left( \frac{n}{R} - \frac{R m}{2} \right) \Omega |n, m \rangle .$
Thus, we have
\begin{equation}
  \Omega |n, m \rangle = |n, - m \rangle,
\end{equation}
which indicates that only highest weight states with winding number $m = 0$ are left-right symmetric.
As a result, the symmetric states contributing to the Klein bottle partition function in Eq.~{\eqref{klein-part2}} can generally be expressed as
\begin{equation}
  |n ; n_1, n_2, n_3, \ldots \rangle = j_{- 1}^{n_1} j_{- 2}^{n_2} \ldots
  \bar{j}_{- 1}^{n_1} \bar{j}_{- 2}^{n_2} \ldots |n, 0 \rangle,
  \label{eq:compbosonCFTsymm}
\end{equation}
where $n_k \geqslant 0, n \in \mathbbm{Z}$.

To evaluate Eq.~{\eqref{klein-part2}}, one can express the zeroth Virasoro generator in terms of $j_k$'s,
\begin{equation}
  L_0 = \frac{1}{2} j_0 j_0 + \sum_{k = 1}^{\infty} j_{- k} j_k,
\end{equation}
and act $L_0$ on $|n ; n_1, n_2, n_3, \ldots \rangle$.
According to the commutation relation $[j_{k_1}, j_{k_2}] = k_1 \delta_{k_1, k_2}$, $[j_{k_1}, \bar{j}_{k_2}] = 0$, we obtain $[j_{- k} j_k, j_{- k}^{n_k}] = k n_k j_{- k}^{n_k}$, and thus $|n ; n_1, n_2, n_3, \ldots \rangle$ is an eigenstate of $L_0$,
\begin{equation}
  L_0 |n ; n_1, n_2, n_3, \ldots \rangle = \left( \frac{1}{2} \frac{n^2}{R^2}
  + \sum_{k = 1}^{\infty} k n_k \right) |n ; n_1, n_2, n_3, \ldots \rangle .
\end{equation}
Therefore, according to Eq.~{\eqref{klein-part2}}, the Klein bottle partition function can be expressed as
\begin{eqnarray}
  Z^{\mathcal{K}} & = & \sum_{n, n_1, n_2, \ldots \in \mathbbm{Z}}
  \langle n ; n_1, n_2, n_3, \ldots |q^{2 (L_0 - c / 24)} |n ; n_1, n_2, n_3,
  \ldots \rangle \nonumber\\
  & = & q^{- c / 12} \sum_{n, n_1, n_2, \ldots \in \mathbbm{Z}}
  q^{\frac{n^2}{R^2} + 2 \sum_{k \geqslant 1} k n_k} \nonumber\\
  & = & q^{- c / 12} \sum_{n \in \mathbbm{Z}} q^{\frac{n^2}{R^2}} \prod_{k =
  1}^{\infty} \frac{1}{1 - q^{2 k}} \nonumber\\
  & = & \theta_3 (2 \tau / R^2) \frac{1}{\eta (2 \tau)},
\end{eqnarray}
where $\eta (\tau) \equiv q^{1 / 24} \prod_{k = 1}^{\infty} (1 - q^k)$ is the Dedekind-$\eta$ function, and $\theta_3 (\tau) \equiv \sum_{n \in \mathbbm{Z}} q^{n^2 / 2}$ is Jacobi's theta function.
To further evaluate $Z^{\mathcal{K}} (\tau)$, one uses the modular transformation of Jacobi's theta and Dedekind-$\eta$ functions,
\begin{eqnarray}
  \sqrt{- \mathrm{i} \tau} \theta_3 (\tau) & = & \theta_3 (- 1 / \tau), \\
  \sqrt{- \mathrm{i} \tau} \eta (\tau) & = & \eta (- 1 / \tau),
\end{eqnarray}
and obtains
\begin{equation}
  Z^{\mathcal{K}} = R \frac{\theta_3 (- R^2 / 2 \tau)}{\eta (- 1 / 2
  \tau)} .
\end{equation}
Under the condition $L \gg v \beta$, we have
\begin{eqnarray}
  \theta_3 \left( \frac{- R^2}{2 \tau} \right) & = & 1 + 2 \mathrm{e}^{- \pi
  \frac{L R^2}{2 \beta v}}, \\
  \eta \left( - \frac{1}{2 \tau} \right) & = & \mathrm{e}^{- \frac{1}{24}
  \frac{\pi L}{\beta v}},
\end{eqnarray}
and therefore
\begin{equation}
  Z^{\mathcal{K}} (L, \beta) = R \mathrm{e}^{\frac{1}{24} \frac{\pi L}{\beta
  v}} .
\end{equation}
When combining with $Z^{\mathcal{T}} (L, \beta) = \mathrm{e}^{\frac{1}{6} \frac{\pi L}{\beta v}}$, we finally arrive at
\begin{equation}
  g = \frac{Z^{\mathcal{K}} (2 L, \beta / 2)}{Z^{\mathcal{T}} (L, \beta)} = R,
\end{equation}
and the Klein bottle entropy is thus
\begin{equation}
  \ln g = \ln R. \label{eq:compactifiedbosongkb}
\end{equation}
Equation {\eqref{eq:compactifiedbosongkb}} is the central result of this work.
When the square of the radius $R^2$ is not a rational number, this result goes beyond the scope of Ref.~{\onlinecite{tu_universal_2017}}, which focuses on rational CFTs. 
Moreover, since the Luttinger liquid corresponds to a compactified boson CFT, and the Luttinger parameter has a direct relationship with the compactification radius, the simple relation Eq.~{\eqref{eq:compactifiedbosongkb}} allows us to determine the Luttinger parameter via computing the Klein bottle entropy of lattice models. 

\subsection{XY chain}

As a concrete example to demonstrate the general result \Eq{eq:compactifiedbosongkb}, we first consider the case of the spin-1/2 XY model
\begin{equation}
  H = - \sum_{i = 1}^L (S_i^x S^x_{i + 1} + S_i^y S_{i + 1}^y).
  \label{eq:XYchain}
\end{equation}
This model is known to be described by a $U (1)_4$ CFT, which is a RCFT and, at the same time, a compactified boson CFT. 
In the meantime, the model also allows exact solution via fermionization. Thus, it provides a nice starting point for checking consistency.  As in the case of the critical Ising chain, we use the lattice reflection operation to interchange the left and right movers of the CFT.

By using the Jordan-Wigner transformation, the model (\ref{eq:XYchain}) is transformed  into a spinless fermion model
\begin{equation}
  H = - \frac{1}{2} \sum_{i = 1}^{L - 1} \left( f_i^{\dagger} f_{i + 1} +
  \mathrm{h.c.} \right) + \frac{1}{2} Q (f_L^{\dagger} f_1 + f_1^{\dagger}
  f_L), \label{eq:XYafterJW}
\end{equation}
where $Q = \mathrm{e}^{\mathrm{i} \pi \sum_{l = 1}^L n_l}$ is the fermion parity.
$Q$ is a conserved quantity and the Hilbert space splits into the Neveu-Schwarz ($Q = 1$ with even number of fermions) and Ramond ($Q = -1$ with odd number of fermions) sectors. In both sectors, the Hamiltonians take the same form
\begin{equation}
  H_{\pm} = - \frac{1}{2} \sum_{i = 1}^L (f_i^{\dagger} f_{i + 1} + f_{i +
  1}^{\dagger} f_i), \label{eq:XYspinlessfermoin}
\end{equation}
with antiperiodic (periodic) boundary condition $f_{L + 1} = - f_1$ ($f_{L + 1} = f_1$) for the Neveu-Schwarz (Ramond) sector. After a Fourier transformation, the Hamiltonian is expressed as
\begin{equation}
  H_{\pm} = - \sum_k \cos k f_k^{\dagger} f_k
  \label{eq:XYspinlessfermionmomentum}
\end{equation}
with the allowed momenta $k = \pm \frac{\pi}{L}, \pm \frac{3 \pi}{L}, \ldots, \pm \frac{(L - 1) \pi}{L}$ in the Neveu-Schwarz sector and $k = 0, \pm \frac{2 \pi}{L}, \ldots, \pm \frac{(L - 2) \pi}{L}, \pi$ in the Ramond sector (we choose $L = 4 m, m \in \mathbbm{N}$ for simplicity). The single-particle energy appearing in (\ref{eq:XYspinlessfermionmomentum}) will be denoted by $E_k = -\cos k$ below.

\subsubsection{Identification of the primary states}

Since the XY model is described by the rational $U(1)_4$ CFT, we start from the perspective of RCFT by identifying all the primary states of the XY chain in the fermion picture and analyzing their behavior under the lattice reflection, as in the case of TFIM.
We also present the energy spectrum of the XY chain of size $L=20$ obtained by means of exact diagonalization calculations, and then identify the primary states obtained from the fermionic picture in the spectrum as a separate check.

\begin{figure}[!htb]
  \resizebox{8.5cm}{!}{\includegraphics{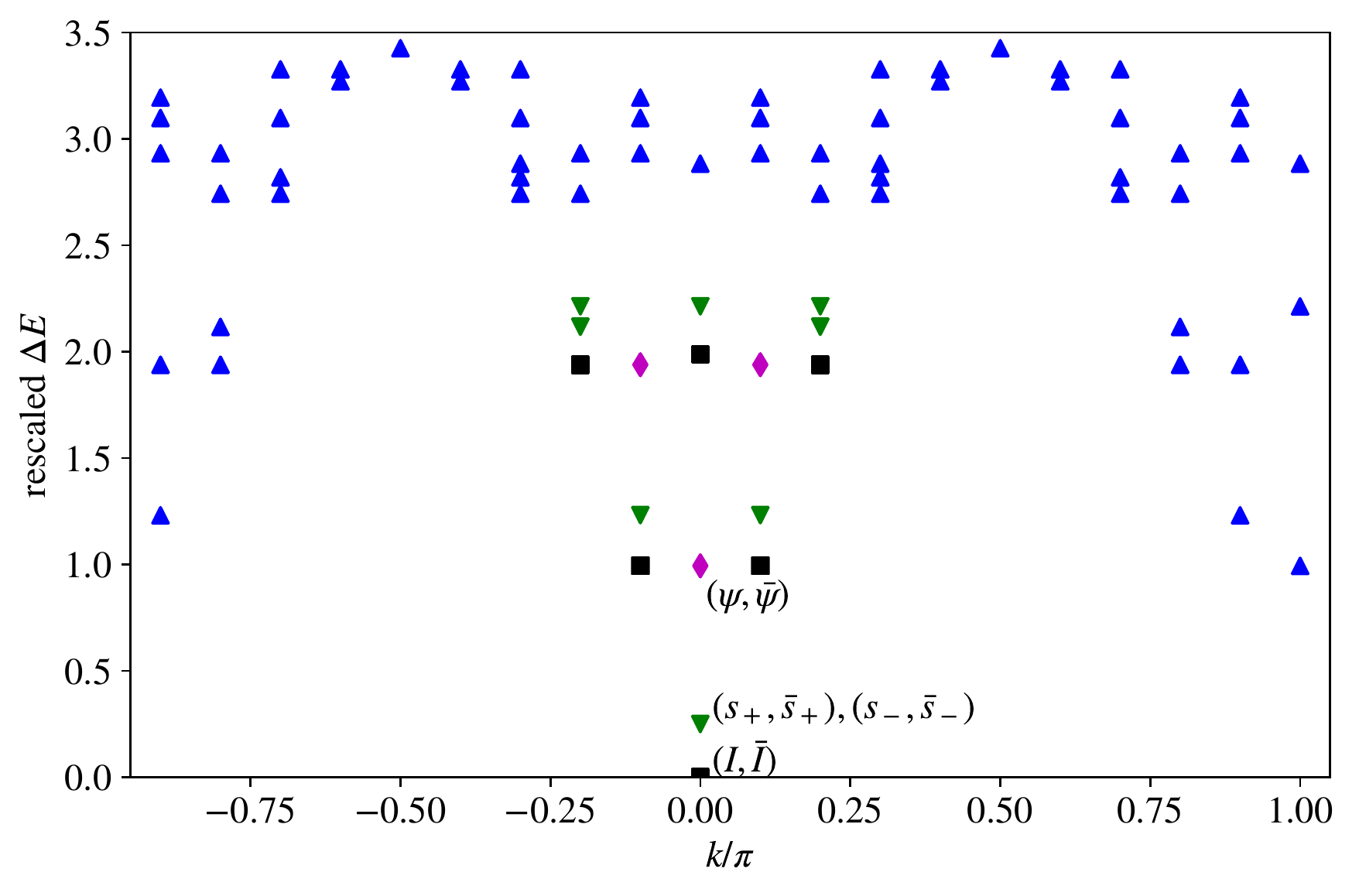}}
  \caption{ The energy spectrum of the XY model with $L = 20$.
  The four primary states are marked by $(I, \bar{I})$, $(s_+, \bar{s}_+)$, $(s_-, \bar{s}_-)$ and $(\psi, \bar{\psi})$ in the spectrum.
  The energy differences $\Delta E$'s have been rescaled according to the scaling dimensions of the primary states.
  The states in the same conformal tower are marked by the same color with the corresponding primary state, while the unidentified states are marked as blue. 
  \label{fig:XYspectrum}}
\end{figure}

The ground state of the system is in the Neveu-Schwarz sector
\begin{equation}
  | \mathrm{gs} \rangle = \prod_{| k | < \pi / 2} f^{\dagger}_k |0 \rangle ,
\end{equation}
and the ground-state energy is
\begin{equation}
  E_{(I, \bar{I})} = - \sum_{| k | < \pi / 2} \cos k = - \frac{1}{\sin
  \frac{\pi}{L}} .
\end{equation}
The ground state has total momentum $k_{\mathrm{tot}} = 0$, and apparently, it is invariant under the lattice reflection. The ground state corresponds to the primary field $(I, \bar{I})$ in the $U (1)_4$ CFT.
This state is labeled by $(I, \bar{I})$ in the energy spectrum (see Fig.~\ref{fig:XYspectrum}).

The lowest excited state in the Neveu-Schwarz sector is obtained by creating a pair of fermions just above the Fermi surface
\begin{equation}
  | \psi, \bar{\psi} \rangle = f_{k = \frac{\pi}{2} + \frac{\pi}{L}}^{\dagger}
  f_{k = - \frac{\pi}{2} - \frac{\pi}{L}}^{\dagger} | \mathrm{gs} \rangle
\end{equation}
with energy
\begin{equation}
  E_{(\psi, \bar{\psi})} = E_{(I, \bar{I})} + 2 \sin \frac{\pi}{L}.
\end{equation}
$| \psi, \bar{\psi} \rangle$ has total momentum
$k_{\ensuremath{\operatorname{tot}}} = 0$, and it is also invariant under the lattice reflection.
This state corresponds to the primary field $(\psi, \bar{\psi})$ in the $U (1)_4$ CFT, which has conformal weight $(h_{\psi}, \bar{h}_{\bar{\psi}}) = (1 / 2, 1 / 2)$.
We label it by $(\psi, \bar{\psi})$ in Fig.~\ref{fig:XYspectrum}.
There is also one degenerate state $f_{k = \frac{\pi}{2} + \frac{\pi}{L}} f_{k = - \frac{\pi}{2} - \frac{\pi}{L}} | \mathrm{gs} \rangle$ which is obtained by annihilating two fermions near the Fermi surface.

In the Ramond sector, the lowest energy states are
\begin{eqnarray}
  |s_+, \bar{s}_+ \rangle & = & \prod_{| k | \leqslant \frac{\pi}{2}}
  f_k^{\dagger} |0 \rangle, \\
  |s_-, \bar{s}_- \rangle & = & \prod_{| k | < \frac{\pi}{2}} f_k^{\dagger} |0
  \rangle .
\end{eqnarray}
These two states are degenerate since $E_{k = \pm\frac{\pi}{2}} = 0$ (note that $k=\pm\frac{\pi}{2}$ are allowed for $L = 4 m, m \in \mathbbm{N}$).
The energy of these two states is
\begin{equation}
  E_{(s_+, \bar{s}_+)} = E_{(s_-, \bar{s}_-)} = - \cot \frac{\pi}{L} .
\end{equation}
Both $|s_+, \bar{s}_+ \rangle$ and $|s_-, \bar{s}_- \rangle$ have total momentum 0 and they are invariant under the lattice reflection.
They correspond to the $U (1)_4$ CFT primary states $(s_+, \bar{s}_+)$ and $(s_-, \bar{s}_-)$ with conformal weight $(h_{s_+}, \bar{h}_{\bar{s}_+}) = (h_{s_-}, \bar{h}_{\bar{s}_-}) = (1 / 8, 1 / 8)$.
These two states are labeled by $(s_+, \bar{s}_+)$ and $(s_-, \bar{s}_-)$ in Fig.~\ref{fig:XYspectrum}.

From Fig.~\ref{fig:XYspectrum}, one may notice that there are also low-energy states with total momentum $\pi$.
According to the discussions in Sec.~\ref{sec:latticereflection}, one may suspect whether these states will contribute to the Klein bottle entropy with a $-1$ factor.
To clarify this, we also identify them in the fermionic picture.
In the Neveu-Schwarz sector, the lowest-energy states
with momentum $\pi$ are $f^{\dagger}_{k = \frac{\pi}{2} + \frac{\pi}{L}} f_{k
= - \frac{\pi}{2} + \frac{\pi}{L}} | \mathrm{gs} \rangle$ and $f_{k =
\frac{\pi}{2} - \frac{\pi}{L}} f^{\dagger}_{k = - \frac{\pi}{2} -
\frac{\pi}{L}} | \mathrm{gs} \rangle$ with energy $E_{(I, \bar{I})} + 2 \sin
\frac{\pi}{L}$.
In the Ramond sector, the lowest-energy states with momentum
$\pi$ are $f^{\dagger}_{k = \frac{\pi}{2} + \frac{2 \pi}{L}} f_{k = -
\frac{\pi}{2} + \frac{2 \pi}{L}} |s_{\pm}, \bar{s}_{\pm} \rangle$ and
$f^{\dagger}_{k = - \frac{\pi}{2} - \frac{2 \pi}{L}} f_{k = \frac{\pi}{2} -
\frac{2 \pi}{L}} |s_{\pm}, \bar{s}_{\pm} \rangle$ with energy $E_{(s_{\pm},
\bar{s}_{\pm})} + 2 \sin \frac{2 \pi}{L}$.
These states are created via the Umklapp process,
and one can easily check that these states have no contribution to the Klein bottle partition function, since they are not left-right symmetric.

From the discussion above, we find that the four primary states of the XY model are all invariant under the lattice reflection operation. These four primary fields are Abelian with quantum dimension $d_a=1$ (total quantum dimension $\mathcal{D}=2$). According to Eq.~{\eqref{eq:gforrationalCFT}}, one has
\begin{equation}
  g_{\mathrm{XY}} = 2, \label{eq:xychainRCFT}
\end{equation}
since $M_{a,a} = 1 \, \forall a$ in the $U(1)_4$ CFT. This result is in agreement with the exact solution of XY model, as shown in Ref.~{\onlinecite{tu_universal_2017}}.
We include the details of the exact solution in Appendix \ref{sec:exacttoXYZs}.

\subsubsection{Identification of the compactification radius}\label{sec:determineR}

From the perspective of the compactified boson CFT, it is crucial to determine the compactification radius to make the prediction on the value of the Klein bottle entropy.
It is well known that there exists a duality in this category of CFTs, which results in the invariance of the torus partition function and the spectrum under the interchange $R \leftrightarrow 2/R$. This duality is called the T duality~\cite{francesco_conformal_1999}.
As indicated by Eq.~\eqref{eq:compactifiedbosongkb}, the T duality is broken on the Klein bottle.
When Eq.~\eqref{eq:compactifiedbosongkb} is applied to lattice models, it cannot be determined in the context of the continuous field theory which radius should be chosen.
Therefore, in lattice models, we need to construct the boson field starting from the microscopic model, and analyze the effect of the lattice reflection in order to determine the compactification radius that should be used.

The low-energy excitations of the XY model are described by a noninteracting Luttinger liquid model, based on which the compactified boson theory is introduced by the bosonization technique.
Following Ref.~{\onlinecite{von_delft_bosonization_1998}}, the Hamiltonian of the Luttinger liquid reads
\begin{equation}
  H = \frac{v_F}{2 \pi} \int_{- L / 2}^{L / 2} \mathrm{d} x : \left[
  \psi^{\dagger}_{\mathrm{L}} (x) \mathrm{i} \partial_x \psi_{\mathrm{L}} (x)
  + \psi_{\mathrm{R}}^{\dagger} (x) (- \mathrm{i} \partial_x)
  \psi_{\mathrm{R}} (x) \right] :, \label{eq:TLLmodelH}
\end{equation}
where $v_F$ is the Fermi velocity.
The normal ordering is defined by $: A := A - \langle A \rangle_{\mathrm{gs}}$, where $\langle A \rangle_{\mathrm{gs}}$ represents the expectation value of the operator $A$ in the ground state $|\mathrm{gs} \rangle$.
The fermion fields $\psi_{\mathrm{L}} (x)$ and $\psi_{\mathrm{R}} (x)$ are defined by $\Psi (x) = \mathrm{e}^{- \mathrm{i} k_F x} \psi_{\mathrm{L}} (x) + \mathrm{e}^{\mathrm{i} k_F x} \psi_{\mathrm{R}} (x)$, where the fermion field $\Psi (x)$ is introduced from the spinless fermion model in Eq.~\eqref{eq:XYafterJW},
\begin{eqnarray}
  \Psi (x) & = & \left( \frac{2 \pi}{L} \right)^{1 / 2} \sum_{p = -
  \infty}^{\infty} \mathrm{e}^{\mathrm{i} p x} f_p \\
  & = & \left( \frac{2 \pi}{L} \right)^{1 / 2} \sum^{\infty}_{k = - \infty}
  \left( \mathrm{e}^{- \mathrm{i} (k_F + k) x} f_{k, \mathrm{L}} +
  \mathrm{e}^{\mathrm{i} (k_F + k) x} f_{k, \mathrm{R}} \right),
  \label{eq:fermionfield2}
\end{eqnarray}
where $f_{k, \mathrm{L} / \mathrm{R}} \equiv f_{\mp (k + k_F)}$.
In the Luttinger liquid, the energy spectrum is linearized, $\varepsilon_{k, \mathrm{L} / \mathrm{R}} = v_F k$, and the range of $k$ has been extended to $(- \infty, + \infty)$, in order to perform the bosonization approach.
The Hamiltonian is then expressed as
\begin{equation}
  H = \sum_{k = - \infty}^{\infty} \sum_{\eta = \mathrm{L,R}} v_F k : f_{k
  \eta}^{\dagger} f_{k \eta} : .
\end{equation}
In terms of the fermion modes $f_{k, \mathrm{L/R}}$, the fermion field $\psi_{\mathrm{L/R}} (x)$ can be written as $\psi_{\mathrm{L/R}} (x) = (2 \pi / L)^{1 / 2} \sum_{k = - \infty}^{+ \infty} \mathrm{e}^{\mp \mathrm{i} k x} f_{k, \mathrm{L/R}}$, and the fermion density $\rho_{\mathrm{L/R}} \equiv : \psi^{\dagger}_{\mathrm{L/R}} \psi_{\mathrm{L/R}} :$ is expressed as
\begin{equation}
  \rho_{\mathrm{L/R}}(x) = \frac{2 \pi}{L} \sum_q \mathrm{e}^{\mp \mathrm{i} q x}
  \sum_k : f_{k - q, \mathrm{L/R}}^{\dagger} f_{k, \mathrm{L/R}} : = \sum_q
  \mathrm{e}^{\mp \mathrm{i} q x} \rho_{q, \mathrm{L/R}},
\end{equation}
where we have introduced $\rho_{q, \mathrm{L/R}} = \frac{2 \pi}{L} \sum_k : f_{k - q, \mathrm{L/R}}^{\dagger} f_{k, \mathrm{L/R}} :$.
For $q = 0$, $\rho_{q, \mathrm{L/R}} = \frac{2 \pi}{L} n_{\mathrm{L/R}}$ corresponds to the number of fermions in the left/right-moving sector, while for $q < 0$ ($q > 0$), $\rho_{q, \mathrm{L/R}}$ creates (annihilate) particle-hole excitations in the corresponding sector.

Under the lattice reflection, according to Eq.~{\eqref{latticePMomentum}}, one can easily check that $P f^{\dagger}_{k, \mathrm{L/R}} P = \mathrm{e}^{\mp \mathrm{i} (L + 1) (k + k_F)} f^{\dagger}_{k, \mathrm{R/L}} Q$, then
\begin{eqnarray}
  P \rho_{q, \mathrm{L/R}} P & = & \frac{2 \pi}{L} \sum_k \left[ P f_{k - q,
  \mathrm{L/R}}^{\dagger} f_{k, \mathrm{L/R}} P - \langle f_{k - q,
  \mathrm{L/R}}^{\dagger} f_{k, \mathrm{L/R}} \rangle_{\mathrm{gs}} \right]
  \nonumber\\
  & = & \mathrm{e}^{\pm \mathrm{i} q (L + 1)} \frac{2 \pi}{L} \sum_k : f_{k -
  q, \mathrm{R/L}}^{\dagger} f_{k, \mathrm{R/L}} : \nonumber\\
  & = & \mathrm{e}^{\pm \mathrm{i} q} \rho_{q, \mathrm{R/L}},
  \label{eq:latticereflectdensity}
\end{eqnarray}
where we have used the fact that the ground state is left-right symmetric and the fermion density $\rho_{\mathrm{L/R}}(x)$ should be periodic in $x$. The modes of the fermion density in the left-moving and right-moving sectors are indeed interchanged under the lattice reflection.
For $q = 0$, the phase factor $\mathrm{e}^{\pm \mathrm{i} q} = 1$, which implies that the numbers of fermions in the left and right sectors are interchanged under the lattice reflection.
For $q \neq 0$, there would be a nonvanishing phase factor.
However, the phase factor would cancel for left-right-symmetric states, while those states that are not symmetric have no contribution to the Klein bottle partition function.

The bosonization method is based on the fact that the particle-hole excitations in one dimension have bosonic nature, due to the commutation relation $[\rho_{q \eta}, \rho_{- q' \eta}] = \frac{2 \pi}{L} q \delta_{q q'} \, (q, q' > 0)$. In fact, one can construct left/right-moving boson field in terms of the particle-hole excitations,
\begin{equation}
  \phi_{\mathrm{L/R}} (x) = \sum_{q \neq 0} \frac{\mathrm{i}}{q} \mathrm{e}^{-
  a q / 2} \mathrm{e}^{\mp \mathrm{i} q x} \rho_{q, \mathrm{L/R}} \pm \rho_{0,
  \mathrm{L/R}} x, \label{eq:notimeboson}
\end{equation}
where $a > 0$ is an infinitesimal regularization parameter to regularize ultraviolet divergent momentum summations (not to be confused with the primary state $a$). The fermion density satisfies
\begin{equation}
  \pm \partial_x \phi_{\mathrm{L/R}} (x) = \rho_{\mathrm{L/R}}(x) .
  \label{eq:fermiondensityandboson}
\end{equation}
and the bosonization identity is
\begin{equation}
  \psi_{\mathrm{L/R}} (x) = a^{- 1 / 2} \mathrm{e}^{\pm \mathrm{i}
  \frac{\pi}{L} x} F_{\mathrm{L/R}} \mathrm{e}^{- \mathrm{i}
  \phi_{\mathrm{L/R}} (x)},
\end{equation}
where $F_{\mathrm{L/R}}$ is the Klein factor, which decreases the fermion number in left-moving (right-moving) branch by one.

To obtain the time dependence of the boson field $\phi_{\mathrm{L/R}}$, one can first write the linearized Hamiltonian in terms of the modes of the particle-hole excitations{~\cite{von_delft_bosonization_1998}},
\begin{equation}
  H = \frac{v_F L}{2 \pi} \left( \sum_{q > 0, \eta = \mathrm{L,R}} \rho_{- q
  \eta} \rho_{q \eta} + \frac{1}{2} \rho_{0 \eta}^2 \right) .
\end{equation}
Using the imaginary-time Heisenberg picture $A (\tau) = \mathrm{e}^{H \tau} A \mathrm{e}^{- H \tau}$, it would be straightforward to obtain that
\begin{eqnarray}
  \rho_{q \eta} (\tau) & = & \rho_{q \eta} \mathrm{e}^{- v_F q \tau}, \\
  F_{\eta} (\tau) & = & F_{\eta} \mathrm{e}^{- v_F \rho_{_{0 \eta}} \tau}
  \mathrm{e}^{\frac{\pi}{L} v_F \tau} .
\end{eqnarray}
Formally, one can absorb the time dependence $\mathrm{e}^{- v_F \rho_{_{0 \eta}} \tau}$ of the Klein factor into the boson field.
The time-dependent boson field then becomes
\begin{equation}
  \phi_{\mathrm{L/R}} (x, \tau) = \sum_{q \neq 0} \frac{\mathrm{i}}{q}
  \mathrm{e}^{- a q / 2} \mathrm{e}^{- q (\pm \mathrm{i} x + v_F \tau)}
  \rho_{q, \mathrm{L/R}} - \mathrm{i} \rho_{0, \mathrm{L/R}} (\pm \mathrm{i} x
  + v_F \tau) . \label{eq:timeboson}
\end{equation}
One can see that $\phi_{\mathrm{L/R}} (x, \tau)$ only depends on $\xi \equiv \mathrm{i} x + v_F \tau$ and $\bar{\xi} \equiv - \mathrm{i} x + v_F \tau$, respectively. The bosonization identity becomes
\begin{equation}
  \psi_{\mathrm{L/R}} (x, \tau) = a^{- 1 / 2} \mathrm{e}^{\frac{\pi}{L} (v_F
  \tau \pm \mathrm{i} x)} F_{\mathrm{L/R}} \mathrm{e}^{- \mathrm{i}
  \phi_{\mathrm{L/R}} (x, \tau)},
\end{equation}
where the Klein factor $F_{\mathrm{L/R}}$ has no time dependence.

Next, from the left/right-moving boson field one can construct a pair of dual fields
\begin{eqnarray}
  \phi (x, \tau) & = & \phi_{\mathrm{L}} (x, \tau) + \phi_{\mathrm{R}} (x,
  \tau),  \label{eq:dualboson1}\\
  \theta (x, \tau) & = & \phi_{\mathrm{L}} (x, \tau) - \phi_{\mathrm{R}} (x,
  \tau) .  \label{eq:dualboson2}
\end{eqnarray}
By writing down $\phi (x, \tau)$ and $\theta (x, \tau)$ explicitly, one can see that the dual boson fields are compactified
bosons{~\cite{francesco_conformal_1999}},
\begin{eqnarray}
  \phi (x, \tau) & = & \frac{2 \pi}{L} \left( n_{\mathrm{L}} - n_{\mathrm{R}}
  \right) x - \frac{2 \pi}{L} \left( n_{\mathrm{L}} + n_{\mathrm{R}} \right)
  (\mathrm{i} v_F \tau) \nonumber\\
  &  & + \frac{\mathrm{i}}{q} \mathrm{e}^{- a q / 2} \sum_{q \neq 0} \left(
  \mathrm{e}^{- q (\mathrm{i} x + v_F \tau)} \rho_{q \mathrm{L}} +
  \mathrm{e}^{- q (- \mathrm{i} x + v_F \tau)} \rho_{q \mathrm{R}} \right),
  \nonumber\\
  &  & \\
  \theta (x, \tau) & = & \frac{2 \pi}{L} \left( n_{\mathrm{L}} +
  n_{\mathrm{R}} \right) x - \frac{2 \pi}{L} \left( n_{\mathrm{L}} -
  n_{\mathrm{R}} \right) (\mathrm{i} v_F \tau) \nonumber\\
  &  & + \frac{\mathrm{i}}{q} \mathrm{e}^{- a q / 2} \sum_{q \neq 0} \left(
  \mathrm{e}^{- q (\mathrm{i} x + v_F \tau)} \rho_{q \mathrm{L}} -
  \mathrm{e}^{- q (- \mathrm{i} x + v_F \tau)} \rho_{q \mathrm{R}} \right),
  \nonumber\\
  &  &
\end{eqnarray}
whose ``zero modes'' $\phi_0$ and $\theta_0$ have already been absorbed into the Klein factor{~\cite{von_delft_bosonization_1998}}.

In the Neveu-Schwarz sector, $n_{\mathrm{L}} + n_{\mathrm{R}} \in 2\mathbbm{Z}$, so $n_{\mathrm{L}} - n_{\mathrm{R}} \in 2\mathbbm{Z}$.
In the Ramond sector, $n_{\mathrm{L}} + n_{\mathrm{R}} \in 2\mathbbm{Z}+ 1$.
However, one needs to note that there exists a fermion mode with zero momentum in the Ramond sector [created by $f^{\dagger}_{k=0}$ in Eq.~\eqref{eq:XYspinlessfermionmomentum}], which is always occupied in the low-energy description and belongs to neither the left-moving nor right-moving sectors.
Formally we can ``split'' this state, and denote $n_{\mathrm{L}}$ and $n_{\mathrm{R}}$ as half-integers, i.e., $n_{\mathrm{L/R}} = n'_{\mathrm{L/R}} + 1 / 2$ with $n_{\mathrm{L/R}}' \in \mathbbm{Z}$. Therefore $n'_{\mathrm{L}} + n_{\mathrm{R}}' \in 2\mathbbm{Z}$ and $n_{\mathrm{L}} - n_{\mathrm{R}} = n'_{\mathrm{L}} - n_{\mathrm{R}}' \in 2\mathbbm{Z}$. As a result, in both sectors $n_{\mathrm{L}} - n_{\mathrm{R}} \in 2\mathbbm{Z}$ and $n_{\mathrm{L}} + n_{\mathrm{R}} \in \mathbbm{Z}$, so $\phi$ has the radius $R = 2$ [note that $\phi (x + L, t) = \phi (x, t) + 2 \pi (n_L-n_R)$] and $\theta$ has the radius $R = 1$ [note that $\theta (x + L, t) = \theta (x, t) + 2 \pi (n_L+n_R)$].
The low-energy physics of the XY model is thus described by two seemingly distinct boson fields with different compactification radius, and the two radii are related by the T duality~\cite{francesco_conformal_1999}.

On the other hand, in the compactified boson CFT, the Laurent modes $j_q$ (or $\bar{j}_q$, the antiholomorphic counterpart) of the $U (1)$ current $j = \mathrm{i} \partial_z \phi$ ($\bar{j} = \mathrm{i} \partial_{\bar{z}} \phi$) of the $R = 2$ boson corresponds to the modes of the fermion density in the left(right)-moving sector, where $z \equiv \mathrm{e}^{2 \pi \xi / L} = \mathrm{e}^{2 \pi (\mathrm{i} x + v_F \tau) / L}$ ($\bar{z} \equiv \mathrm{e}^{2 \pi \bar{\xi} / L} = \mathrm{e}^{2 \pi (-\mathrm{i} x + v_F \tau) / L}$).
Comparing Eqs.~{\eqref{eq:notimeboson}}, \eqref{eq:fermiondensityandboson} and \eqref{eq:timeboson}, and noting $\partial_z = \frac{1}{z} \frac{L}{2 \pi} \partial_{\xi}$ and $\partial_{\bar{z}} = \frac{1}{\bar{z}} \frac{L}{2 \pi} \partial_{\bar{\xi}}$, one can see that
\begin{equation}
  j_q = \frac{L}{2 \pi} \rho_{q, \mathrm{L}}, \; \bar{j}_q = \frac{L}{2 \pi}
  \rho_{q, \mathrm{R}} . \label{eq:jqrouqcorresp}
\end{equation}
Meanwhile, for the field $\theta$ with radius $R = 1$,
\begin{equation}
  j'_q = \frac{L}{2 \pi} \rho_{q, \mathrm{L}}, \; \bar{j}'_q = - \frac{L}{2 \pi}
  \rho_{q, \mathrm{R}},
\end{equation}
where $j'_q = \mathrm{i} \partial_z \theta$ and $\bar{j}' = \mathrm{i} \partial_{\bar{z}} \theta$.
According to Eq.~{\eqref{eq:latticereflectdensity}} and the discussions below, for the field $\phi$ with $R=2$, the lattice reflection $P$ indeed interchanges the holomorphic and the antiholomorphic sectors, up to a factor that will cancel in the symmetric states.
In contrast, for its dual field $\theta$ with $R=1$, the lattice reflection will introduce an additional minus sign.
Therefore, one can see that the bond-centered lattice reflection in the XXZ model matches the reflection operation in the CFT Hilbert space of the field $\phi$ with $R = 2$, instead of its dual field $\theta$. Using Eq.~{\eqref{eq:compactifiedbosongkb}}, we get
\begin{equation}
  \ln g = \ln R = \ln 2.
\end{equation}
This is consistent with the RCFT result in Eq.~\eqref{eq:xychainRCFT}.

What is more, from the above discussion, one can gain a physical interpretation of the states in Eq.~{\eqref{eq:compbosonCFTsymm}} that contribute the Klein bottle entropy.
From Eq.~{\eqref{eq:jqrouqcorresp}}, the highest weight states $|n, m \rangle$ are annihilated by any $j_q$ $(q > 0)$ that annihilates the particle-hole excitations, so $|n, m \rangle$ represents the Fermi-sea states with $n, m \in \mathbbm{Z}$ representing the number of fermions, correspondingly $\frac{n}{2} + m$ and $\frac{n}{2} - m$ in the left-moving and right-moving sector.
Therefore, $j_{- 1}^{n_1} j_{- 2}^{n_2} \ldots \bar{j}_{- 1}^{n_1} \bar{j}_{- 2}^{n_2} \ldots |n, 0 \rangle$ represents the state with the same number of fermions and same particle-hole excitations in the two sectors, which is apparently left-right symmetric and thus makes a contribution to the Klein bottle entropy.

\subsection{XXZ model}\label{sec:introduceXXZ}

Next, we consider the spin-1/2 XXZ model by adding a nearest-neighboring Ising interaction to the XY chain,
\begin{equation}
  H = - \sum_{i = 1}^L (S_i^x S^x_{i + 1} + S_i^y S_{i + 1}^y) + \Delta
  \sum_{i = 1}^L S_i^z S_{i + 1}^z, \label{eq:xxzhamiltonian}
\end{equation}
where $\Delta$ is an anisotropy coefficient.
For $- 1 < \Delta \leqslant 1$, the system is in the Luttinger liquid phase, and its low-energy physics can be described by a compactified boson CFT{~\cite{giamarchi_quantum_2003}}. 
For general value of $\Delta$ within this phase, CFT prediction on the Klein bottle entropy is the first result which goes beyond the RCFT results of Ref.~\onlinecite{tu_universal_2017}. 

As in the case of XY model, the XXZ model can be transformed into a spinless fermion model by Jordan-Wigner transformation, with an additional interaction term $H_{\mathrm{int}}$ compared to the XY model, i.e., $H = H_{\mathrm{0}} + H_{\mathrm{int}}$ with $H_{\mathrm{0}}$ representing the noninteracting fermion model obtained from the XY model, and $H_{\mathrm{int}}$ reads
\begin{equation}
  H_{\mathrm{int}} = \Delta \sum_{i = 1}^L \left( f_i^{\dagger} f_i -
  \frac{1}{2} \right) \left( f_{i + 1}^{\dagger} f_{i + 1} - \frac{1}{2}
  \right) . \label{eq:interactioninXXZ}
\end{equation}
Since the fermion parity $Q$ is still conserved, we can again split the Hilbert space into the Neveu-Schwarz and Ramond sectors with different fermion parities $Q = \pm 1$ and boundary conditions $f_{L + 1} = \mp f_1$. In the frame of bosonization one can obtain the underlying compacitified boson CFT of this system, which leads to the CFT prediction of the Klein bottle entropy in this model.

\subsubsection{CFT prediction}

To obtain the compacitified boson description of the XXZ model, we pass to the continuum limit.
The interaction term $H_{\mathrm{int}}$ can be written as{~\cite{affleck_field_1989}}
\begin{equation}
  H_{\mathrm{int}} = H_{\mathrm{d-d}} + H_{\mathrm{Umklapp}},
\end{equation}
where we have introduced the local fermion-fermion interaction term $H_{\mathrm{d-d}}$ and the Umklapp term $H_{\mathrm{Umklapp}}$ that scatter the fermion between different sectors
\begin{eqnarray}
  H_{\mathrm{d-d}} & = & \Delta \int_{- L / 2}^{L / 2} \frac{\mathrm{d} x}{2
  \pi} : \left( \rho_{\mathrm{L}}^2 + \rho_{\mathrm{R}}^2 + 4
  \rho_{\mathrm{L}} \rho_{\mathrm{R}} \right) :, \\
  H_{\mathrm{Umklapp}} & = & - 2 \Delta \int_{- L / 2}^{L / 2}
  \frac{\mathrm{d} x}{2 \pi} : \left[ \left( \psi_{\mathrm{L}}^{\dagger}
  \psi_{\mathrm{R}} \right)^2 + \mathrm{h.c.} \right] : .
\end{eqnarray}
In the Luttinger liquid phase, by renormalization group analysis, the Umklapp process $H_{\mathrm{Umklapp}}$ is irrelevant for $- 1 < \Delta < 1$ and marginally irrelevant at the Heisenberg point $\Delta = 1$, while for $\Delta > 1$ $H_{\mathrm{Umklapp}}$ becomes relevant and introduces a mass term which causes the system to be gapped~\cite{giamarchi_quantum_2003}.

For $- 1 < \Delta \leqslant 1$, the interaction generally renormalizes the parameters and the interaction term $H_{\mathrm{d-d}}$ can be written as
\begin{equation}
  H_{\mathrm{d-d}} = \int_{- L / 2}^{L / 2} \frac{\mathrm{d} x}{2 \pi} :
  \left[ \frac{1}{2} g_4 \left( \rho_{\mathrm{L}}^2 + \rho_{\mathrm{R}}^2
  \right) + g_2 \rho_{\mathrm{L}} \rho_{\mathrm{R}} \right] :,
\end{equation}
where $g_2$ and $g_4$ are undetermined coefficients which depend on the specific choice of the parameter $\Delta$. Correspondingly, the kinetic term $H_0$ can be represented in terms of the fermion densities,
\begin{equation}
  H_0 = \frac{v_F}{2 \pi} \int_{- L / 2}^{L / 2} \mathrm{d} x : \frac{1}{2}
  \left( \rho_{\mathrm{L}}^2 + \rho_{\mathrm{R}}^2 \right) :,
\end{equation}
and the total Hamiltonian $H = H_{\mathrm{0}} + H_{\mathrm{d-d}}$ can be written in a diagonalized form as
\begin{equation}
  H = \frac{v}{4} \int_{- L / 2}^{L / 2} \frac{\mathrm{d} x}{2 \pi} : \left[
  \frac{1}{K} \left( \rho_{\mathrm{L}} + \rho_{\mathrm{R}} \right)^2 + K
  \left( \rho_{\mathrm{L}} - \rho_{\mathrm{R}} \right)^2 \right] :,
\end{equation}
where $v = \sqrt{(v_F + g_4)^2 - g_2^2}$ is the velocity and $K = \sqrt{\frac{v_F + g_4 - g_2}{v_F + g_4 + g_2}}$ is called the Luttinger parameter, which is usually used as a parametrization of the interaction strength in the system.
Generally, the values of $v$ and $K$ cannot be reliably obtained from field theory calculations and one has to resort to the microscopic models to determine their values.
In the case of the spin-1/2 XXZ model, $v$ and $K$ are determined by the Bethe ansatz solution{~\cite{giamarchi_quantum_2003}}
\begin{eqnarray}
  K & = & \frac{\pi}{2 (\pi - \cos^{- 1} \Delta)}, \\
  v & = & \frac{\pi}{2} \frac{\sqrt{1 - \Delta^2}}{\cos^{- 1} \Delta} .
\end{eqnarray}
According to {\eqref{eq:fermiondensityandboson}}, {\eqref{eq:dualboson1}} and {\eqref{eq:dualboson2}}, the Hamiltonian can be expressed in terms of the boson fields,
\begin{equation}
  H = \frac{v}{4} \int_{- L / 2}^{L / 2} \frac{\mathrm{d} x}{2 \pi} : \left[
  \frac{1}{K} (\partial_x \theta)^2 + K (\partial_x \phi)^2 \right] : .
\end{equation}
Note that in the noninteracting case (XY limit), $g_2 = g_4 = 0$, and $v = K = 1$.
As a result, the interaction effectively rescales the compactified bosons as $\Theta = \frac{\theta}{\sqrt{K}}, \Phi = \sqrt{K} \phi$, and the Hamiltonian becomes
\begin{equation}
  H = \frac{v}{4} \int_{- L / 2}^{L / 2} \frac{\mathrm{d} x}{2 \pi} :
  [(\partial_x \Theta)^2 + (\partial_x \Phi)^2] :,
\end{equation}
where $\Theta$ has radius $1 / \sqrt{K}$ and $\Phi$ has radius $2 \sqrt{K}$.
According to the discussion in the case of the XY model in Sec.~\ref{sec:determineR}, one concludes that the Klein bottle entropy should be calculated based on the boson field $\Phi$, which gives
\begin{equation}
  g = 2 \sqrt{K} = \sqrt{\frac{2 \pi}{\pi - \cos^{- 1} \Delta}} . \label{eq:CFTofxxzspinhalf}
\end{equation}

\subsubsection{Numerical results} \label{subsubsec:numericalResultofXXZ}

Next we verify Eq.~\eqref{eq:CFTofxxzspinhalf} numerically.
We employ a quantum Monte Carlo simulation in the XXZ chain, by calculating the partition function ratio $g = Z^{\mathcal{K}} (2 L, \beta / 2) / Z^{\mathcal{T}} (L, \beta)$ using an improved version of the extended ensemble Monte Carlo method{~\cite{tang_universal_2017}}.
We include the details of the algorithm in Appendix \ref{sec:extendedMCmethod}.

\begin{figure}[!htb]
  \resizebox{8.5cm}{!}{\includegraphics{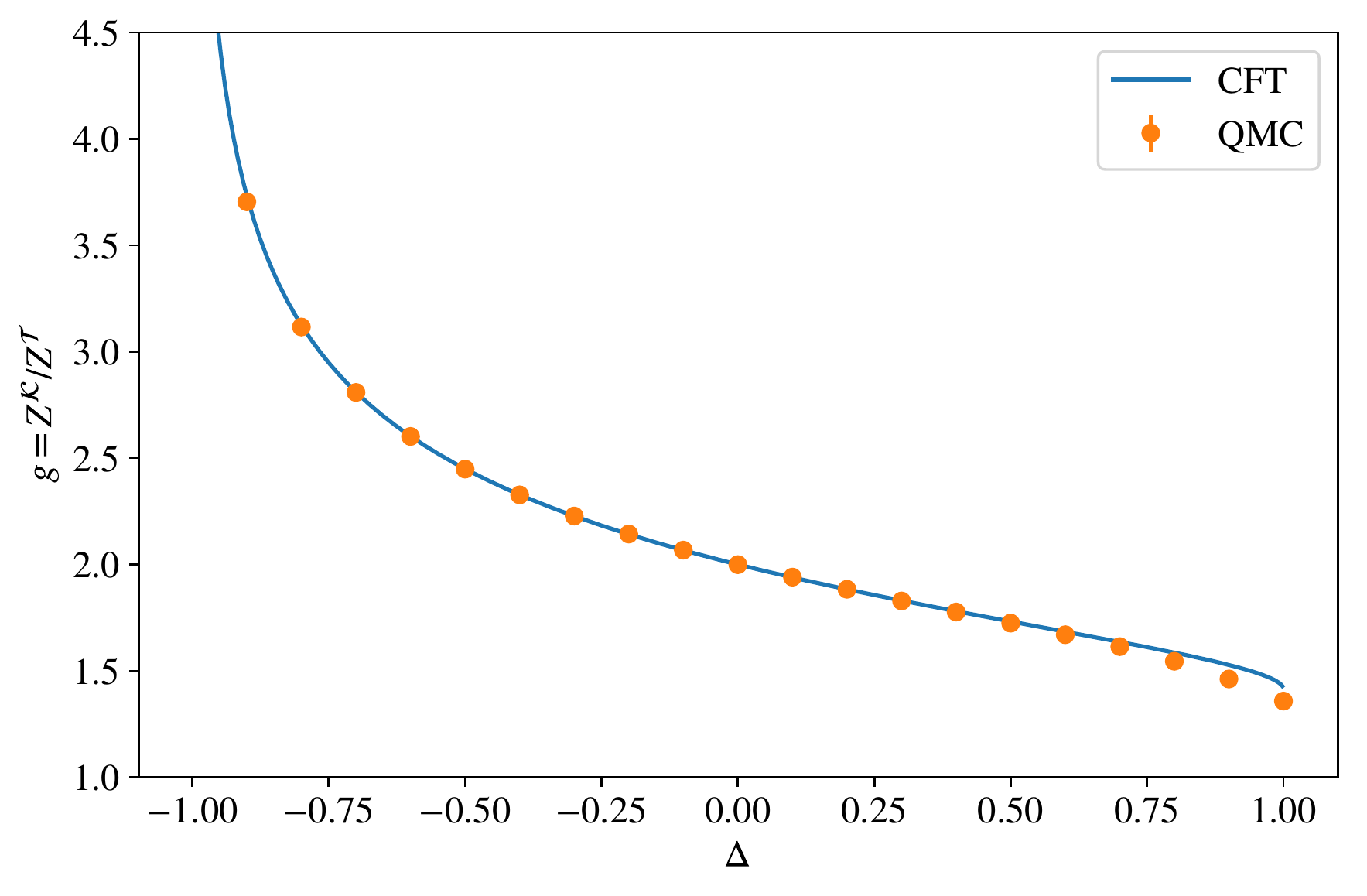}}
  \caption{ Comparison of the QMC result of the Klein bottle entropy with the CFT prediction.
  The error bars are smaller than the data points.
  In the QMC calculation, the parameters are chosen as $L = 440, \beta = 44$, where $g$ is calculated by $Z^{\mathcal{K}} (2 L, \beta / 2) / Z^{\mathcal{T}} (L, \beta)$, according to Eq.~{\eqref{eq:Kleinbottleexpression}}. \label{fig:XXZgkb}}
\end{figure}

As shown in Fig.~\ref{fig:XXZgkb}, one can see that the numerical results and the CFT predictions are in good agreement with each other, except in the vicinity of $\Delta = 1$.
The slight deviation may originate from the marginally irrelevant Umklapp process at $\Delta = 1$.
We have observed this kind of slight deviation in the $q = 4$ Potts model, which also has a marginally irrelevant term{~\cite{tang_universal_2017}}.

The remarkable agreement of the CFT prediction and the QMC numerical results indicates that the Klein bottle entropy can be a reliable tool to extract the Luttinger parameter $K$ from the lattice models. 
Comparing to the existing methods{~\cite{giamarchi_quantum_2003,song_general_2010,song_bipartite_2012,dalmonte_critical_2012,dalmonte_estimating_2011,lauchli_operator_2013,alcaraz_in_preparation_2018}},
the advantage of the present approach is that one can directly obtain the Luttinger parameter by calculating the Klein bottle entropy in a finite temperature calculation, without any fitting procedure.

\subsection{Spin-1 XXZ model}

Next we employ our QMC method to the more challenging spin-1 XXZ model,
where the Hamiltonian still takes the form of Eq.~{\eqref{eq:xxzhamiltonian}},
but the operators $S^{\nu} (\nu = x, y, z)$ are now spin-1 operators.
We calculate the Klein bottle entropy in this model for $- 1 < \Delta \leqslant 1$.
The spin-1 XXZ model is in the Luttinger liquid phase only in the range $- 1 < \Delta \leqslant 0$. While for $0 < \Delta \leqslant 1$, the system is in the massive Haldane phase with a finite energy gap{~\cite{botet_ground-state_1983,alcaraz_critical_1992,kitazawa_phase_1996}}.

\begin{figure}[!htb]
  \resizebox{8.5cm}{!}{\includegraphics{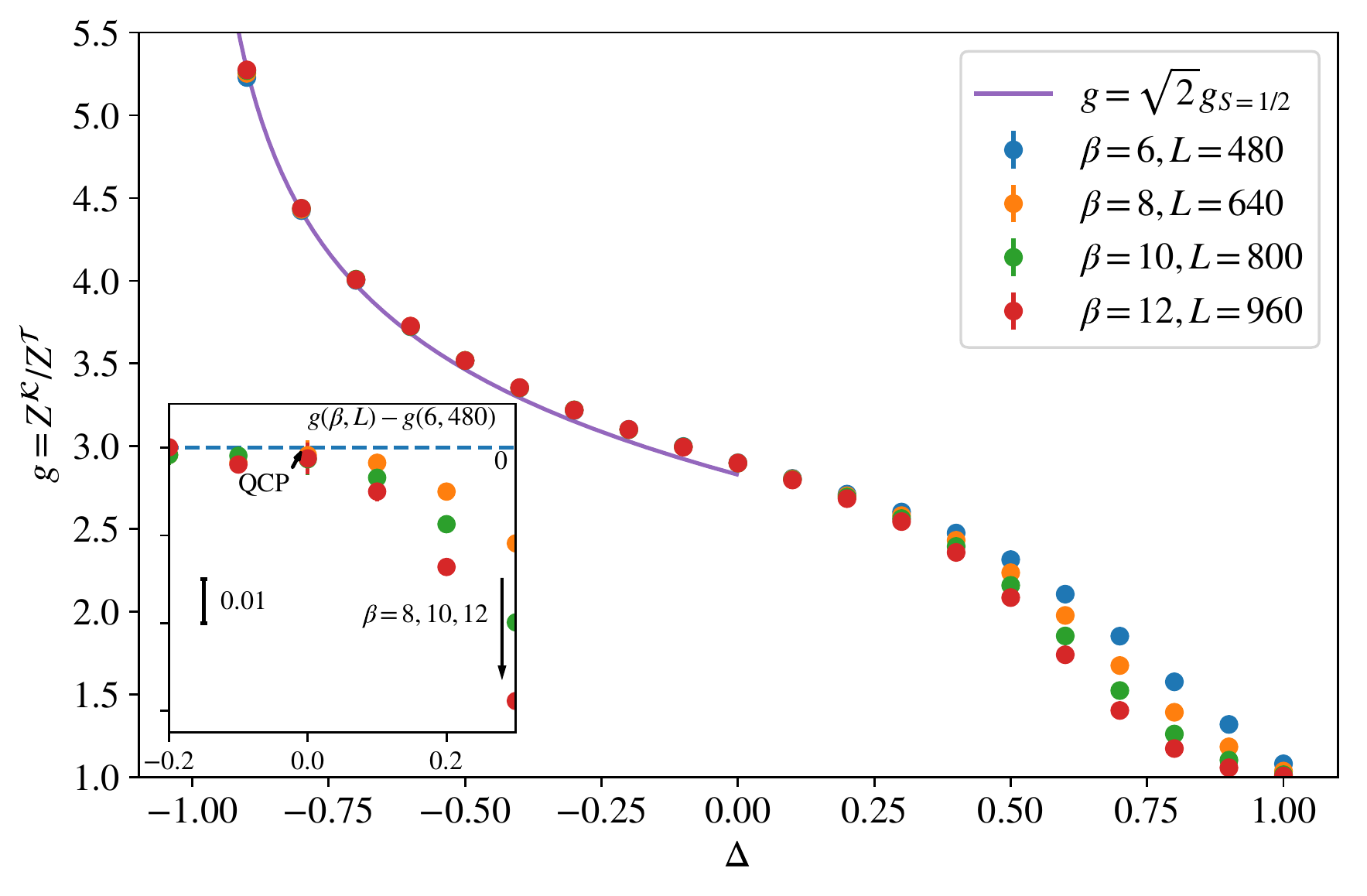}}
  \caption{ The QMC result for the Klein bottle entropy for $- 1 < \Delta \leqslant 1$.
  The error bars are smaller than the data points.
  For $1 < \Delta \leqslant 0$, the system is in the Luttinger liquid phase, and the Klein bottle degeneracy $g$ gives the compactification radius $R$.
  The solid line is the conjectured Luttinger liquid parameter for $S = 1$ in Ref.~\onlinecite{alcaraz_critical_1992}.
  For $\Delta > 0$, the system is in the gapped Haldane phase, and the Klein bottle entropy varies with the different temperature $\beta$ and system size $L$, in contrast to the case of critical phase.
  The deviation of the Klein bottle entropy of different parameters starts at the quantum critical point $\Delta = 0$, as shown in the inset, where $\Delta g = g (\beta, L) - g (\beta = 6, L = 480)$ is plotted. \label{fig-spin1-gkb}}
\end{figure}

In contrast to the case of $S=1/2$, the spin-1 XXZ model cannot be exactly solved. 
According to the relation $g=R=2\sqrt{K}$, our numerical results of the Klein bottle entropy can be used to conversely determine the Luttinger parameter $K$.
We can compare our numerical result with the conjecture proposed in Ref.~{\onlinecite{alcaraz_critical_1992}} for the Luttinger parameter in the spin-$S$ XXZ model, $K_S = 2 S K_{S = 1 / 2}$, which is equivalent to
\begin{equation}
  g_S = \sqrt{2 S} g_{S = 1 / 2}. \label{eq:spin1conjecture}
\end{equation}
For $S = 1$, we have $g_{S = 1} = \sqrt{2} g_{S = 1 / 2}$.
As shown in Fig.~\ref{fig-spin1-gkb}, the conjectured formula is in good agreement with the numerical results up to some small deviations.

The conjecture \eqref{eq:spin1conjecture} was proposed based on the finite-size-scaling results of the exact diagonalization data~\cite{alcaraz_critical_1992}, and currently there is no rigorous proof for this conjecture.
However, based on the symmetry analysis, one can show that the XY ($\Delta = 0$) point of this conjecture is exact.
On the one hand, in the context of the continuous field theory, it is known that there is an inherent SU(2) symmetry in the Berezinskii-Kosterlitz-Thouless (BKT) transition point~\cite{halpern_quantum_1975,halpern_equivalent-boson_1976,banks_bosonization_1976,ginsparg_curiosities_1988,nomura_su2_1998}.
At the BKT transition point, the Luttinger parameter is restricted, and possible choices include $K=1/2$ (corresponding to the SU(2)$_1$ Wess-Zumino-Witten model) and $K=2$~\cite{nomura_su2_1998}.
On the other hand, a hidden SU(2) symmetry was found in the spin-1 XY model~\cite{kitazawa_su_2003}.
Together with the exact diagonalization results given by Ref.~{\onlinecite{alcaraz_critical_1992}}, which indicate $K=2$ in the spin-1 XY model,
one can infer that the BKT transition between the Luttinger-liquid phase and the Haldane phase locates exactly at the XY point ($\Delta = 0$),
and this point precisely corresponds to $g=R=2\sqrt{2}$.
As one can see in Fig.~\ref{fig-spin1-gkb}, there exists some small deviation between our numerical result and the exact result at the XY point, which is again attributed to the marginally irrelevant term, since the BKT transition is driven by the marginal operator.

We also calculate the Klein bottle entropy out of the critical region into the gapped Haldane phase.
With $\Delta$ passing the BKT transition point $\Delta = 0$, the originally degenerate values of the Klein bottle entropy at different $\beta$ and $L$ start to deviate with each other, as shown in Fig.~\ref{fig-spin1-gkb}.
The deviation starts exactly at the BKT transition point between the two phases~\cite{chen_conformal_2017}.
We note that it was a difficult task to determine the BKT transition point from numerical calculations, due to the exponentially small energy gap towards the BKT transition point in the gapped phase~\cite{kosterlitz_ordering_1973, kosterlitz_critical_1974}.
In the Haldane phase, the value of the Klein bottle entropy will eventually converge to the ground-state degeneracy of the system on the Klein bottle when $\beta$ is large enough.

\subsection{The Affleck-Ludwig entropy}

As a comparison, we also attempted to extract the compactification radius by calculating the Affleck-Ludwig (AL) entropy in the spin-1 XXZ model{~\cite{affleck_universal_1991}}.
The AL entropy emerges from the open boundary of a long cylinder, which is universal and only depends on the CFT and conformal boundary conditions.
In lattice models, when $L \gg v \beta$, one can obtain the AL entropy by calculating the ratio between the partition functions of the systems on a long cylinder and a torus{~\cite{tang_universal_2017}},
\begin{equation}
  \ln \left( \frac{Z^{\mathcal{C}}}{Z^{\mathcal{T}}} \right) \approx
  S_{\ensuremath{\operatorname{AL}}} - f_b \beta, \label{eq:ALasZratio}
\end{equation}
where $f_b$ is the surface free energy density, which is a nonuniversal quantity. By a linear extrapolation, one can get the AL entropy as the intercept.

In the compactified boson CFT, the AL entropy is also dependent on the compactification radius $R$.
For simplicity, we only consider the case that the boundary conditions on the two boundaries are the same.
For the Dirichlet and Neumann boundary condition, we have{~\cite{oshikawa_boundary_2010}}
\begin{eqnarray}
  S_{\mathrm{AL}}^{\mathrm{D}} & = & \ln (R / 2), \\
  S_{\mathrm{AL}}^{\mathrm{N}} & = & - \ln R.
\end{eqnarray}
In the XXZ model, the fixed (free) boundary condition of spin chain corresponds to the Neumann (Dirichlet) boundary condition for the free boson{~\cite{eggert_magnetic_1992,affleck_edge_1998}}. 
For simplicity, we only performed the QMC calculations for the free boundary condition.
However, in practice, due to the existence of the nonuniversal term $- f_b \beta$, the partition function ratio decays exponentially with $\beta$, and the error of the calculation becomes intolerable when $\beta$ reaches some certain value.
On the other hand, the calculation result from the finite-size lattice will converge to the universal value only when the $\beta$ and $L$ is large enough{~\cite{tang_universal_2017}}.
In our calculations of the spin-1 XXZ model, unfortunately, the range of $\beta$ where we are able to perform the calculation cannot give the accurate value of AL entropy.
The difficulty here highlights the advantage of using Klein bottle entropy, which is free of nonuniversal surface energies and does not need any extrapolation procedure.

\section{Summary}\label{sec:summary}

To summarize, in this paper, we first review the results and details of the initial work Ref.~{\onlinecite{tu_universal_2017}} which focuses on the Klein bottle entropy in RCFT and discuss in detail how to extract the Klein bottle entropy from lattice model calculations via the bond-centered lattice reflection.
We then go beyond the scope of RCFT and study the Klein bottle entropy in the compactified boson CFT, which contains both rational and nonrational CFTs.
We obtain a simple relation between the Klein bottle entropy and the compactification radius, $\ln g = \ln R$, which is the central result of our work.
Due to the direct connection between the compactification radius and the Luttinger parameter, our result provides a straightforward and efficient method to extract the Luttinger parameter from lattice models.

In lattice models, we employ quantum Monte Carlo calculations in the XXZ chain with $S = 1 / 2$ and $S = 1$, respectively.
For the exactly solvable spin-$1 / 2$ XXZ chain, our numerical results show excellent agreement with the CFT prediction, except the slight deviations near the isotropic point $\Delta = 1$, which we attribute to the marginally irrelevant fields.
For the $S = 1$ XXZ chain that cannot be exactly solved, our numerical results serve as a new numerical determination of the Luttinger parameter in this model.

\section*{Acknowledgment}
We thank F.~C.~Alcaraz and G.~Sierra for helpful discussions.
This work is supported by NSF-China under Grant No.11504008 (W.T. and X.C.X),
Ministry of Science and Technology of China under the Grant No.2016YFA0302400 (L.W.)
and the DFG via project A06 of SFB 1143 (H.H.T.).
The simulation is performed at Tianhe-1A platform at the National Supercomputer Center in Tianjin.
Parts of the calculations are performed using the ALPS library{~\cite{albuquerque_alps_2007,bauer_alps_2011}}.

\clearpage
\appendix
\section{Improved extended ensemble Monte Carlo method for partition function ratios}\label{sec:extendedMCmethod}

\subsection{General description}

To compute the partition function ratios using Monte Carlo methods, one can use the extended ensemble Monte Carlo method.
The basic idea is that, in order to obtain the partition function ratio $Z^{\mathcal{A}} / Z^{\mathcal{B}}$ where $Z^{\eta}$ corresponds to the system put on the manifold $\eta = \mathcal{A}, \mathcal{B}$,
we can perform an extended-ensemble simulation whose partition function is written as
\begin{equation}
  Z = Z^{\mathcal{A}} + Z^{\mathcal{B}} = \sum_{\eta \in \{ \mathcal{A}, \mathcal{B} \} } \sum_C W^{\eta} (C),
\end{equation}
where $W^{\eta} (C)$ represents the Boltzmann weight of the configuration $C$ in the ensemble $\eta$.
During the simulation, we treat the configuration $C$ and the label $\eta$ on equal footing, i.e., the Monte Carlo updates also
include transitions between ensembles that update the label $\eta$~\cite{tang_universal_2017}.

A typical issue of such methods is that the acceptance rate of the transition between different ensembles usually decays exponentially with the system size.
Usually, one can overcome this issue by introducing some intermediate systems and compute the partition function as
\begin{equation}
  \frac{Z^{\mathcal{A}}}{Z^{\mathcal{B}}} = \frac{Z^{\mathcal{A}}}{Z^{(1)}}
  \frac{Z^{(1)}}{Z^{(2)}} \ldots \frac{Z^{(m)}}{Z^{\mathcal{B}}},
\end{equation}
where $Z^{(n)}, n = 1, 2, \ldots, m$ is the partition function of the intermediate systems.

Here we present another possible improvement for this method in combination with loop/cluster update employed in our simulation~\cite{todo_loop_2013, gubernatis_quantum_2016}.
Similar tricks have been applied to the Swendsen-Wang algorithm in classical systems{~\cite{caraglio_entanglement_2008,alba_entanglement_2013}} and to stochastic series expansion (SSE) in quantum spin systems{~\cite{kulchytskyy_detecting_2015}},
which, together with the loop/cluster algorithm, all share the same framework of ``two-step selection''{~\cite{gubernatis_quantum_2016}}.

Generally, in the loop/cluster algorithm of path-integral QMC, the update procedure consists of two steps{~\cite{gubernatis_quantum_2016}}. In the first step, we stochastically generate a graph $G$ from the current configuration $C$ with probability $P (G|C)$,
and in the second step, we generate the new configuration $C'$ from the graph $G$ with probability $P (C' |G)$.
Therefore, during the simulation, there exists another graph space $\Gamma$ other than the configuration space $\Sigma$.
By introducing a new weight $W (C, G)$ defined by $W (C) = \sum_{G \in \Gamma} W (C, G)$ in the joint space $\Sigma \times \Gamma$,
one can write the probabilities $P (G|C)$ and $P (C' |G)$ as
\begin{equation}
  P (G|C) = \frac{W (C, G)}{W (C)}, P (C' |G) = \frac{W (C', G)}{W (C)} .
\end{equation}
One can then define the Boltzmann weight of the graph as
\begin{equation}
  W (G) \equiv \sum_{C \in \Sigma} W (C, G) . \label{eq:WGoriginal}
\end{equation}

In the most straightforward manner of the extended-ensemble method,
one proposes a transition from ensemble $\mathcal{A}$ to the other ensemble $\mathcal{B}$ directly in the configuration space without modifying the configuration $C$,
and calculate the acceptance ratio as
\begin{equation}
  r (C ; \mathcal{A} \rightarrow \mathcal{B}) = \min \left( 1,
  \frac{W^{\mathcal{B}} (C)}{W^{\mathcal{A}} (C)} \right) .
\end{equation}
Usually, $W^{\mathcal{B}} (C) / W^{\mathcal{A}} (C) = \mathrm{e}^{-\beta (E^{\mathcal{B}} (C) - E^{\mathcal{A}} (C))} \neq 1$
since the same configuration $C$ usually has different energies in different ensembles.
The energy difference $E^{\mathcal{B}} (C) - E^{\mathcal{A}} (C)$ often scales with the system size $L$ or the temperature $\beta$.
As a result, the acceptance rate of the transitions between ensembles usually decays exponentially with $L$ or $\beta$.

As a concrete example, one can consider the Ising model at the critical point, whose spins usually form large domains.
When the system configuration is put onto another manifold, the original domains of spins slip and mismatch,
and this usually results in an increase of the system energy.
Qualitatively speaking, such increase of the energy usually scales with the system size,
which will lead to the exponential decay of the acceptance rate of the transitions.
Another example is the XXZ chain discussed in the present paper,
where the spin configuration in the path-integral formulation forms closed loops due to conservation of the $S_z$ magnetization.
If one directly put the configuration onto another path-integral manifold, these closed loops often gets broken, which leads to an illegal configuration, and the transition update will be rejected.
One in general anticipates that the chances of the closed loops not getting cut will decay exponentially with the system size.

The improvement we present here is to propose the transition between ensembles in the graph space $\Gamma$ instead of the configuration space $\Sigma$.
After we generate the graph $G$ from the configuration $C$, we can propose the transition to the other ensemble based on the graph $G$.
Usually, the graph $G$ cannot be directly put in the other ensemble, since the lattice sites have different connection relations on different manifolds.
However, the graph elements in the graph $G$ can be manipulated with much more freedom than the spins in the configuration $C$\footnote{For the models that don't have a freezing problem, all possible graphs are compatible with the worldline configurations, and therefore the graph elements can be manipulated with absolute freedom. For those models with a freezing problem, however, there exist graphs that are incompatible with worldline configurations ({see Ref.~\onlinecite{todo_loop_2013}}). In these cases, the loop algorithm cannot be used, and one may need to resort to the worm algorithm, which is out of the scope of our discussion.}.
This allows us to manipulate the graph elements in the original graph $G$ according to the topology of the targeted manifold.
After the manipulation, a new graph $G'$ is generated, and from this new graph, we can then generate the new configuration $C'$, which naturally resides on the targeted manifold.
The acceptance ratio of this graph manipulation operation can be determined by the Boltzmann weight of the graphs $G$ and $G'$,
\begin{equation}
  r (G \rightarrow G' ; \mathcal{A} \rightarrow \mathcal{B}) = \min \left( 1,
  \frac{W^{\mathcal{B}} (G')}{W^{\mathcal{A}} (G)} \right) ,
\end{equation}
where $W^{\eta} (G)$ represents the Boltzmann weight of the graph $G$ in the ensemble labeled by $\eta= \mathcal{A}, \mathcal{B}$. 

Next, we derive the acceptance ratio in the cluster/loop algorithm, following the conventions of
Ref.~{\onlinecite{gubernatis_quantum_2016}}. 
During the update procedure, when generating the configuration $C$ from the graph $G$, if we flip the clusters with even probability, then
\begin{equation}
  P (C|G) = \frac{W (C, G)}{W (G)} = \frac{1}{q^M}, \label{wcgwg}
\end{equation}
where $q$ is the number of spin states (for example, for the spin-$1/2$ system, $q = 2$), $M$ is the number of clusters in the graph.
On the other hand, in the loop/cluster algorithm, the joint weight $W (C, G)$ can be expressed as{~\cite{gubernatis_quantum_2016}}
\begin{equation}
  W (C, G) = \prod_p \omega_p (C^p, G^p) = \prod_p v_p (G^p) \Delta (C^p,
  G^p), \label{wcg}
\end{equation}
where $p$ is the ``plaquette'', $\Delta (C^p, G^p)$ is 1 if $C^p$ and $G^p$ are compatible and equals 0 otherwise.
$v_p (G^p)$ is the Boltzmann weights of the graph elements, which depends only on the type of the graph element.
Now we specify a configuration $C_0$ that is compatible with graph $G$.
Then in Eq.~{\eqref{wcg}} all $\Delta (C_0^p, G^p) = 1$. Therefore
\begin{equation}
  W (C_0, G) = \prod_p v_p (G^p) . \label{wc0g}
\end{equation}
According to Eqs.~{\eqref{wcgwg}}{\eqref{wc0g}},
\begin{equation}
  W (G) = \frac{W (G)}{W (C_0, G)} W (C_0, G) = q^M \prod_p v_p (G^p) .
  \label{wg}
\end{equation}
When manipulating the graph elements in $G$, if we only move the locations of the graph elements without removing any of them or adding new ones, the number of graph elements of each type is unchanged.
Then the jump acceptance ratio only depends on the number of clusters formed in the graphs $G$ and $G'$,
\begin{equation}
  r (\mathcal{A} \rightarrow \mathcal{B}) = \min \left( 1,
  \frac{W^{\mathcal{B}} (G')}{W^{\mathcal{A}} (G)} \right) = \min (q^{M_{G'} -
  M_G}, 1) , \label{acra}
\end{equation}
where $M_G$ represents the number of clusters in the graph $G$.

In practice, one can usually improve the acceptance rate by a large factor by performing the transition update in the graph space, since this trick actually expands the overlap between the two ensembles.
As an example, in the XXZ chain where we perform our simulation, if we propose the transition update directly in the configuration space, most of such updates will be rejected, since there are usually graphs locating between the boundary sites, which will lead to the cut of closed loops in the worldline configuration after the direct transition. 
In other words, the direct transition update, which is originally developed for the quantum Potts model~\cite{tang_universal_2017}, will usually break the total spin conservation which is satisfied by the XXZ model.
In this case, the overlap between the two ensembles only consists of those configurations with no graph elements between the boundaries, which will quickly decay with respect to the system size $L$ and the temperature $\beta$. 
As a comparison, our new method allows us to propose the ensemble transition update at any possible configuration, i.e., the overlap between the two ensembles is expanded to the whole configuration space, which will greatly improve the efficiency of the sampling process.

\subsection{Klein bottle entropy}

\begin{figure}[!htb]
  \resizebox{8.5cm}{!}{\includegraphics{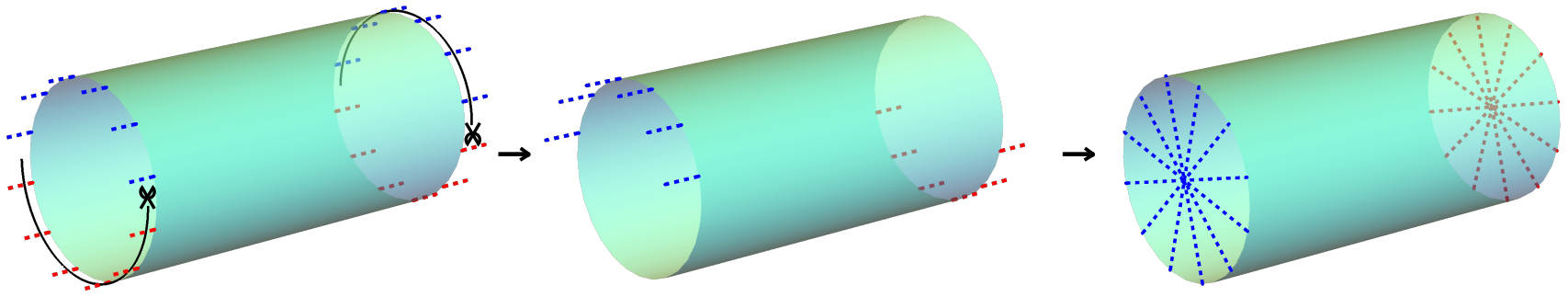}}
  \caption{ The manipulation of the graph elements during the transition from the torus ensemble to the Klein bottle ensemble.
  The graph elements on the boundaries are separated into two groups, which are marked by blue and red dashed lines, correspondingly.
  In practice, one can separate these graph elements by, for example, putting the graph elements within the range $(0, \beta/2)$ into one group and those within the range $(\beta/2, \beta)$ into another group.
  For the blue graph elements, we cut their connections with the right boundary, and connect them to the opposite location on the left boundary. Similar operations are performed for the red graph elements.
  After the manipulation, the graph elements on the boundaries are apparently compatible with the long-range interactions on the cylinder boundary, which correspond to the Klein bottle manifold.
  Conversely, to propose a transition from the Klein bottle ensemble to the torus ensemble, one only need to reverse the manipulation procedure described above.
 \label{fig:manipulation-gelem}}
\end{figure}

\begin{figure*}[!htb]
  \resizebox{17.5cm}{!}{\includegraphics{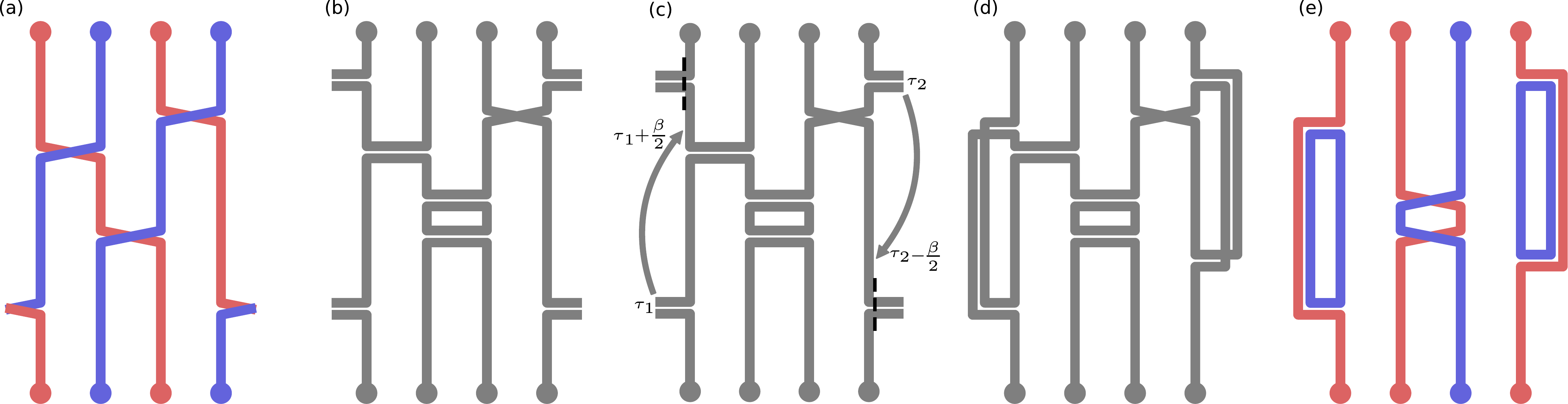}}
  \caption{The steps within one Monte Carlo update.
  Here we use a system with size $L=4$ and inverse temperature $\beta$ as an example.
  (a) We start with a worldline configuration in the torus ensemble.
  The different colors of the worldlines represent different spin orientations.
  The worldlines form closed loops, and this configuration cannot be directly put in the Klein bottle ensemble.
  (b) From the worldline configuration, one can generate a graph $G$ by inserting graph elements into the worldline configuration.
  The graph elements form $M_{G} = 3$ clusters.
  (c) To propose a transition to the Klein bottle ensemble, one has to manipulate the graph elements on the two boundaries according to the connection relations in the latter manifold.
  For the graph element at $\tau_1$ in the lower half $(0, \beta/2)$ of the imaginary time axis, we cut its connection with the right boundary and connect it to the location $\tau_1 + \beta/2$ at the left boundary.
  Meanwhile, for the graph element at $\tau_2$ in the upper half $(\beta/2, \beta)$, we do the opposite.
  One may note that there exist different choices in the cutting and reconnection procedure described above, i.e., for a graph element locating between the boundary sites, one can cut its connection with either the left or the right boundary site.
  To avoid this ambiguity, which may cause a problem in the detailed balance condition, one only needs to fix the choice for all Monte Carlo steps. 
  For example, for the graph in the range $(0, \beta/2)$, we always cut its connection with the right boundary, and for the graph in the range $(\beta/2, \beta)$, we always choose to cut its connection with the left boundary. 
  Reversely, when performing the transition from the Klein bottle ensemble to the torus ensemble, for the graph on the left boundary, we always cut its connection in the range $(\beta/2, \beta)$, and for the graph on the right boundary, we always cut its connection in the range $(0, \beta/2)$.
  (d) The new graph $G'$ that resides on the Klein bottle.
  There are $M_{G'} = 2$ clusters in this graph.
  The acceptance ratio according to \Eq{acra} is $r = \min (1, 2^{M_{G'} - M_{G}}) = 0.5$.
  If this update is accepted, we can proceed to randomly flip the clusters in this graph, and obtain a new worldline configuration in the Klein bottle ensemble.
  Otherwise, one needs to go back to the original graph in (b) and generate new worldline configurations in the torus ensemble based on it.
  (e) The new worldline configuration in the Klein bottle ensemble, generated from the graph in (d). \label{fig:TransformOfWorldline}}
\end{figure*}

Next, we discuss the details of the calculation of the Klein bottle entropy of the XXZ chain using the improved extended ensemble method.
In numerical calculations, we obtain the Klein bottle entropy by calculating the partition function ratio
\begin{equation}
g = Z^{\mathcal{K}} (2 L,\beta / 2) / Z^{\mathcal{T}} (L, \beta).
\end{equation}
During the simulation, we use the standard loop algorithm~\cite{todo_loop_2013, gubernatis_quantum_2016}, and the two manifolds in the two ensembles of the extended ensemble simulation are correspondingly a torus and a Klein bottle.
Although the Klein bottle and the torus have different sizes along the spatial and imaginary-time directions, as shown in Ref.~{\onlinecite{tang_universal_2017}}, one can transform the Klein bottle with parameters $\beta / 2$ and $2 L$ into a cylinder with parameters $\beta$ and $L$ that has long-range interactions on the boundaries.
The cylinder and the torus have the same size along both the spatial and imaginary-time directions, and their differences only locate in the boundary conditions.
Therefore, the transition between the two ensembles only involves the transformations on the boundaries.
In our improved extended ensemble simulation, we only need to manipulate the graph elements on the boundaries according to the topology of the targeted manifold, as shown in Fig.~\ref{fig:manipulation-gelem}.
To be more specific, for the spin-1/2 XXZ chain, we show the details of a single Monte Carlo update in Fig.~\ref{fig:TransformOfWorldline}, using the language of the worldlines in the loop algorithm.

For the XXZ chain with $S=1$, one can decompose each spin operator into the sum of two spin-$1/2$ operators, and apply a projection operator in order to project the expanded Hilbert space onto the original Hilbert space{~\cite{kawashima_loop_1994,gubernatis_quantum_2016}}. Its partition function in the torus ensemble is expressed as
\begin{equation}
  Z = \mathrm{Tr} (\mathcal{P} \mathrm{e}^{-\beta \tilde{H}}),
\end{equation}
where $\tilde{H}$ represents the Hamiltonian of the spin-1/2 system obtained from the decomposition, and $\mathcal{P} = \sum_i \mathcal{P}_i = \frac{1}{2} \sum_{\pi,i} D_i (\pi)$ represents the projection operator, with $\mathcal{P}_i$ representing the projection on each site, and $D_i (\pi)$ representing the permutation $\pi : \{ 1, 2 \} \rightarrow \{ \pi(1), \pi (2)\}$.
We note that the projection operator commutes with both the lattice reflection $P$ and the spin-1/2 Hamiltonian $\tilde{H}$.
Therefore, the extended ensemble Monte Carlo update procedure is the same as that of $S=1/2$, despite the more complicated lattice structure.

\subsection{Affleck-Ludwig entropy}\label{sec:ALQMC}

The AL entropy can be obtained by calculating the partition function ratio
$Z^{\mathcal{C}} (L,\beta) / Z^{\mathcal{T}} (L, \beta)$
and performing a linear extrapolation along $\beta$, according to Eq.~\eqref{eq:ALasZratio}.
Here $\mathcal{C}$ represents the cylinder, due to the open boundary condition.
To calculate the partition function ratio above, we again use the method of extended ensemble simulation and still perform the transition operations in the graph space.
Here we only consider the case of free boundary condition.
We note that there exist plenty of graphs with nonidentical graph elements between the site $1$ and site $L$ in the torus ensemble, and these graphs are all forbidden in the cylinder ensemble.
In contrast, if a graph has no nonidentical graph elements on the boundaries, it can reside in both ensembles, and its Boltzmann weights in the two ensembles are related by a multiplicative constant, which originates from the additional identical graph elements between site $1$ and site $L$ on the torus{~\cite{gubernatis_quantum_2016}},
\begin{equation}
  W^{\mathcal{C}} (G) / W^{\mathcal{T}} (G) = (1 + a \Delta \tau)^{- \beta /
  \Delta \tau} = \mathrm{e}^{- a \beta},
\end{equation}
where $a$ is a constant, and we have taken the continuous limit $\Delta \tau \rightarrow 0$.

We handle the multiplicative constant by a reweighting procedure.
During the extended ensemble simulation, we set $r (\mathcal{T} \rightarrow \mathcal{C})= n (L, 1)$
and $r (\mathcal{C} \rightarrow \mathcal{T}) = 1$,
where $n (i, j)$ equals one if there are no nonidentical graph elements between site $i$ and site $j$, and equals zero otherwise.
Therefore, the partition function ratio that we obtain from the extended ensemble calculation is actually $\mathrm{e}^{a \beta} Z^{\mathcal{C}} (G) / Z^{\mathcal{T}} (G)$.
The prefactor $\mathrm{e}^{a \beta}$ won't affect the result of the AL entropy since it can be absorbed into the nonuniversal surface energy.

In practice, since $r (\mathcal{C} \rightarrow \mathcal{T}) = 1$, as indicated in Ref.~{\onlinecite{broecker_renyi_2014}},
we can conveniently obtain the partition function ratio by performing the simulation only in the torus ensemble
\begin{equation}
  \frac{Z^{\mathcal{C}}}{Z^{\mathcal{T}}} = \frac{\langle r (\mathcal{T}
  \rightarrow \mathcal{C}) \rangle_{\mathcal{T}}}{\langle r (\mathcal{C}
  \rightarrow \mathcal{T}) \rangle_{\mathcal{C}}} = \langle n (L, 1)
  \rangle_{\mathcal{T}} .
\end{equation}
We improve this estimator by
\begin{equation}
  \frac{Z^{\mathcal{C}}}{Z^{\mathcal{T}}} = \frac{1}{L} \left \langle \sum_{i = 1}^L
  n (i, i + 1) \right\rangle_{\mathcal{T}} = \frac{\langle N_{\mathrm{empty}}
  \rangle_{\mathcal{T}}}{L},
\end{equation}
where we have identified site $L + 1$ with site $1$ and introduced the quantity $N_{\mathrm{empty}} \equiv \sum_{i = 1}^L n (i, i + 1)$ that represents the number of intervals that are ``empty'', i.e., contain no nonidentical graph elements.

\section{Exact solution to the partition functions for critical Ising chain and XY chain}\label{sec:JWsolTFIMXY}

\subsection{The critical Ising chain}\label{subsec:exacttoising}

For the critical Ising chain (\ref{eq:IsingHamil}), the torus partition function is given by $Z^{\mathcal{T}} = \mathrm{Tr} (\mathrm{e}^{-
\beta H}) = \mathrm{Tr}_{\mathrm{NS}} (\mathrm{e}^{- \beta H_+})
+\mathrm{Tr}_{\mathrm{R}} (\mathrm{e}^{- \beta H_-})$,
where
$\mathrm{Tr}_{\mathrm{NS}}$ and $\mathrm{Tr}_{\mathrm{R}}$
represent the trace over the Neveu-Schwarz and the Ramond sector, respectively.
In the Neveu-Schwarz sector, we represent the states by $\gamma^\dagger_{k_1} \gamma^\dagger_{k_2} \ldots \gamma^\dagger_{k_N} | \mathrm{gs} \rangle$ with $k_1 < k_2 < \ldots < k_N$ belonging to the allowed lattice momenta $K_{\mathrm{NS}} = \left\{ \pm \frac{\pi}{L}, \pm \frac{3 \pi}{L}, \ldots, \pm \frac{(L - 1) \pi}{L} \right\}$. 
To enumerate these states, we introduce a set of numbers $\mathcal{F}=\{ F_k \}$ ($k\in K_{\mathrm{NS}}$) for each state, where $F_k = 1$ if the fermion mode $\gamma^\dagger_k$ is occupied, and $F_k = 0$ otherwise.
Since the fermion parity is even in the Neveu-Schwarz sector, we have
\begin{equation}
  \mathrm{Tr}_{\mathrm{NS}} (\mathrm{e}^{- \beta H_+}) = \mathrm{e}^{-\beta E_{\mathrm{gs}}} \sum_{\mathcal{F}} \prod_{k \in K_{\mathrm{NS}}} \mathrm{e}^{- \beta F_k E_k} \frac{1 + (- 1)^{N}}{2},
\end{equation}
where $E_{\mathrm{gs}}$ is the ground-state energy of the critical Ising chain [see Eq.~{\eqref{eq:isinggsE}}], and $E_k = \cos (k/2)$. Here we have introduced a factor $\frac{1 + (- 1)^{N}}{2}$ to remove the states with odd fermion parity $Q=-1$. Note that $N = \sum_{k \in K_{\mathrm{NS}}}F_k$, the summation over $\mathcal{F}=\{F_k\}$ can then be carried out for each $F_k=\pm 1$ individually,
\begin{equation}
  \mathrm{Tr}_{\mathrm{NS}}
(\mathrm{e}^{- \beta H_+})
= \frac{1}{2} \mathrm{e}^{- \beta E_{\mathrm{gs}}} \left[ \prod_{k \in K_{\mathrm{NS}}} (1 + \mathrm{e}^{- \beta E_k}) + \prod_{k \in K_{\mathrm{NS}}} (1 - \mathrm{e}^{- \beta E_k}) \right].
\end{equation}

In the Ramond sector, similarly, we represent the states by $\gamma^\dagger_{k_1} \gamma^\dagger_{k_2} \ldots \gamma^\dagger_{k_N} | \sigma, \bar{\sigma} \rangle$ with $k_1 < k_2 < \ldots < k_N$ belonging to the allowed lattice momenta $K_{\mathrm{R}} = \left\{ 0, \pm \frac{2 \pi}{L}, \pm \frac{4 \pi}{L}, \ldots, \pm \frac{(L - 2) \pi}{L}, \pi \right\}$, where we have defined $\gamma^\dagger_0 \equiv f_0$ and $\gamma^\dagger_\pi \equiv f^\dagger_\pi$ for convenience.
By again introducing a set of numbers $\mathcal{F}=\{ F_k \}$ ($k\in K_{\mathrm{R}}$) for each state, one can similarly obtain
\begin{equation}
  \mathrm{Tr}_{\mathrm{R}} (\mathrm{e}^{- \beta H_-}) =
  \frac{1}{2} \mathrm{e}^{- \beta E_{(\sigma,\bar{\sigma})}} \left[ \prod_{k \in
  K_{\mathrm{R}}} (1 + \mathrm{e}^{- \beta E_k}) + \prod_{k \in
  K_{\mathrm{R}}} (1 - \mathrm{e}^{- \beta E_k}) \right],
  \label{eq:TrRamondTFIM}
\end{equation}
where $E_{(\sigma,\bar{\sigma})}$ is the energy of the first excited state of the critical Ising chain [see Eq.~{\eqref{eq:isingsigmaE}}] and $E_k = \cos (k/2)$ (note that we have absorbed $E_{k=\pi} = 0$ and $E_{k=0} = 1$ in this expression, which correspond to $\gamma^\dagger_\pi = f^\dagger_\pi$ and $\gamma^\dagger_0 = f_0$,  respectively).
Since $E_{k=\pi}=0$, the second term in the square bracket of Eq.~\eqref{eq:TrRamondTFIM} vanishes, and the torus partition function reads
\begin{eqnarray}
  Z^{\mathcal{T}} (L, \beta) & = & \frac{1}{2} \mathrm{e}^{- \beta
  E_{\mathrm{gs}}} \left[ \prod_{k \in K_{\mathrm{NS}}} (1
  + \mathrm{e}^{- \beta E_k}) + \prod_{k \in K_{\mathrm{NS}}} (1 -
  \mathrm{e}^{- \beta E_k}) \right] \nonumber\\
  &  & + \frac{1}{2} \mathrm{e}^{- \beta E_{(\sigma,\bar{\sigma})} } \prod_{k \in
  K_{\mathrm{R}}} (1 + \mathrm{e}^{- \beta E_k}) .  \label{eq:ZTexact}
\end{eqnarray}

For the Klein bottle partition function (obtained by the bond-centered lattice reflection, see Sec.~\ref{sec:latticereflection}), the fermion modes all group in pairs (except $k = 0, \pi$ in the Ramond sector). In the Neveu-Schwarz sector, one has
\begin{equation}
  \mathrm{Tr}_{\mathrm{NS}} (P \mathrm{e}^{- \beta H_+}) = \mathrm{e}^{- \beta
  E_{\mathrm{gs}}} \prod_{k \in K_{\mathrm{NS}}, k > 0} (1
  + \mathrm{e}^{- 2 \beta E_k}),
\end{equation}
while for the Ramond sector, recall that states built from $| (\sigma, \bar{\sigma}) \rangle$ have parity $1$, and states built from $f^{\dagger}_{k = \pi} f_{k = 0} |\sigma,\bar{\sigma}\rangle$ have parity $- 1$, one then obtains
\begin{equation}
  \mathrm{Tr}_{\mathrm{R}} (P \mathrm{e}^{- \beta H_-}) = \mathrm{e}^{- \beta
  E_{(\sigma,\bar{\sigma})} } (1 - \mathrm{e}^{- \beta (E_{k = \pi} + E_{k = 0})}) \prod_{k
  \in K_{\mathrm{R}}, 0 < k < \pi} (1 + \mathrm{e}^{- 2 \beta E_k}) .
\end{equation}
Therefore, the Klein bottle partition function for the Ising chain is given by
\begin{eqnarray}
  Z^{\mathcal{K}} (L, \beta) & = & \mathrm{e}^{- \beta
  E_{\mathrm{gs}}} \prod_{k \in K_{\mathrm{NS}}, k > 0} (1
  + \mathrm{e}^{- 2 \beta E_k}) \nonumber\\
  &  & + \mathrm{e}^{- \beta E_{(\sigma,\bar{\sigma})}} (1 - \mathrm{e}^{- \beta})
  \prod_{k \in K_{\mathrm{R}}, 0 < k < \pi} (1 + \mathrm{e}^{- 2 \beta E_k}) .
  \label{eq:exacttoisingZK}
\end{eqnarray}
One can use the exact solutions of the partition functions Eqs.~{\eqref{eq:ZTexact}} and {\eqref{eq:exacttoisingZK}} to calculate the Klein bottle entropy $\ln g$ by Eq.~{\eqref{eq:Kleinbottleexpression}}, under the condition $L \gg v \beta$, where the velocity is $v=1/2$ in the Ising chain.

\subsection{XY chain}\label{sec:exacttoXYZs}

For the XY chain (\ref{eq:XYchain}), the method for calculating the partition functions is very similar to the Ising case in Appendix \ref{subsec:exacttoising}. The torus partition function takes a form similar to Eq.~{\eqref{eq:ZTexact}}, 
\begin{eqnarray}
  Z^{\mathcal{T}} (L, \beta) & = & \frac{1}{2}  \prod_{k \in K_{\mathrm{NS}}}
  (1 + \mathrm{e}^{- \beta E_k}) + \frac{1}{2}  \prod_{k \in K_{\mathrm{NS}}}
  (1 - \mathrm{e}^{- \beta E_k}) \nonumber\\
  &  & + \frac{1}{2}  \prod_{k \in K_{\mathrm{R}}} (1 + \mathrm{e}^{- \beta
  E_k}),  \label{eq:XYtorusexact}
\end{eqnarray}
where $E_k = - \cos k$.
On the other hand, similarly as Eq.~{\eqref{eq:exacttoisingZK}}, the Klein bottle partition function reads
\begin{eqnarray}
  Z^{\mathcal{K}} (L, \beta) & = & \prod_{k \in K_{\mathrm{NS}}, k > 0} (1 +
  \mathrm{e}^{- 2 \beta E_k}) \nonumber\\
  &  & + (\mathrm{e}^{\beta} - \mathrm{e}^{- \beta}) \prod_{k \in
  K_{\mathrm{R}}, 0 < k < \pi} (1 + \mathrm{e}^{- 2 \beta E_k}) .
  \label{eq:XYkleinexact}
\end{eqnarray}
Again, Eqs.~{\eqref{eq:XYtorusexact}} and {\eqref{eq:XYkleinexact}} can be used to calculate the Klein bottle entropy $\ln g$ by Eq.~{\eqref{eq:Kleinbottleexpression}}, under the condition $L \gg v \beta$, where the Fermi velocity is $v=1$ in the XY chain.


\bibliography{xxzKlein}
\bibliographystyle{apsrev4-1}

\end{document}